\documentclass[12pt]{article}

\usepackage{verbatim,color,amssymb}
\usepackage{amsmath}					
\usepackage{amsthm}					
\usepackage{algorithm,algorithmic}
\usepackage{natbib}
\usepackage{setspace}
\usepackage[mathscr]{euscript}
\usepackage{fancyhdr}
\usepackage{enumitem}
\usepackage{graphicx}
\usepackage{geometry}
\usepackage{lineno}
\usepackage[compact]{titlesec}
\usepackage{listings}

\usepackage{subfig}
\usepackage{caption}
\usepackage{float}
\usepackage{wrapfig}
\newsubfloat{figure}
\usepackage{tikz}
\usetikzlibrary{arrows,chains,backgrounds,fit}
\usepackage{multirow}
\usepackage{lineno}

\newtheorem{Thm}{\underline{\bf Theorem}}

\newtheorem*{Proof*}{Proof}

\newtheorem{Lem}{\underline{\bf Lemma}}

\def\etal{\emph{et al.}}

\def\C{{\cal C}}

\def\calP{{\cal P}}
\def\S{{\cal S}}

\def\Y{{\cal Y}}

\def\wh{\widehat}
\def\wt{\widetilde}

\def\log{\hbox{log}}

\def\Beta{\hbox{Beta}}

\def\Dir{\hbox{Dir}}

\def\Unif{\hbox{Unif}}
\def\Mult{\hbox{Mult}}

\def\JASA{{\it Journal of the American Statistical Association}}

\def\P_25_ICML{{\it Proceedings of the 25th international conference on Machine learning}}

\def\refhg{\hangindent=20pt\hangafter=1}
\def\refmark{\par\vskip 2mm\noindent\refhg}

\def\refhg{\hangindent=20pt\hangafter=1}
\def\refmark{\par\vskip 2mm\noindent\refhg}

\def\bse{\begin{eqnarray*}}
\def\ese{\end{eqnarray*}}
\def\be{\begin{eqnarray}}
\def\ee{\end{eqnarray}}
\def\bq{\begin{equation}}
\def\eq{\end{equation}}

\def\wh{\widehat}

\def\trans{^{\rm T}}

\def\th{^{th}}

\def\b1e{{\mathbf e}}

\def\bG{{\mathbf G}}

\def\bk{{\mathbf k}}
\def\bM{{\mathbf M}}

\def\bs{{\mathbf s}}

\def\bU{{\mathbf U}}

\def\bw{{\mathbf w}}

\def\by{{\mathbf y}}

\def\bz{{\mathbf z}}

\newcommand{\bpi}{\mbox{\boldmath $\pi$}}

\newcommand{\bzeta}{\mbox{\boldmath $\zeta$}}

\newcommand{\blambda}{\mbox{\boldmath $\lambda$}}

\renewcommand\footnoterule{\kern-3pt \hrule \textwidth 2in \kern 2.6pt}

\def\boxit#1{\vbox{\hrule\hbox{\vrule\kern6pt \vbox{\kern6pt \textcolor{blue}{#1}\kern6pt}\kern6pt\vrule}\hrule}}

\def\authorfootnote#1{{\let\thefootnote\relax\footnotetext{#1}}}

\allowdisplaybreaks

\begin{document}
\thispagestyle{empty}
\baselineskip=28pt

\begin{center}
{\LARGE{\bf Bayesian Nonparametric Modeling of Higher Order Markov Chains}}
\end{center}
\baselineskip=12pt

\begin{center}
Abhra Sarkar and David B. Dunson\\
Department of Statistical Science, Duke University, Box 90251, Durham NC 27708-0251\\
abhra.sarkar@duke.edu and dunson@duke.edu\\
\end{center}

\vskip 8mm
\begin{center}
{\Large{\bf Abstract}} 
\end{center}
\baselineskip=12pt
We consider the problem of flexible modeling of higher order Markov chains when an upper bound on the order of the chain is known but the true order and nature of the serial dependence are unknown. 
We propose Bayesian nonparametric methodology based on conditional tensor factorizations, which  
can characterize any transition probability with a specified maximal order.  
The methodology selects the important lags and captures higher order interactions among the lags, while also facilitating calculation of Bayes factors for a variety of hypotheses of interest.  
We design efficient Markov chain Monte Carlo algorithms for posterior computation, allowing for uncertainty in the set of important lags to be included and in the nature and order of the serial dependence. 
The methods are illustrated using simulation experiments and real world applications.

\vskip 8mm
\baselineskip=12pt
\noindent\underline{\bf Some Key Words}: Bayesian nonparametrics, Categorical time series, Conditional tensor factorization, Higher order Markov chains, Sequential categorical data. 

\par\medskip\noindent
\underline{\bf Short Title}: Higher Order Markov Chains

\par\medskip\noindent

\pagenumbering{arabic}
\setcounter{page}{0}
\newlength{\gnat}
\setlength{\gnat}{16.5pt}
\baselineskip=\gnat

\newpage

\section{Introduction} \label{sec: introduction}
For $t=1,\ldots,T$, consider a time indexed sequence of categorical variables $\{y_{t}\}$. 
We assume that the distribution of $y_{t}$ may depend on the values at the previous $q$ time points, $y_{t-1},\ldots,y_{t-q}$. 
For $t=(q+1),\dots,T$, the transition probability law governing the evolution of the sequence satisfies 
\bse
p(y_{t} \mid y_{t-1},\dots,y_{1}) = p(y_{t} \mid y_{t-1},\dots,y_{t-q}),  
\ese
and the likelihood function of the sequence admits the factorization
\bse
p(\by_{1:T}) = p_{0}(\by_{1:q})\prod_{t=(q+1)}^{T}p(y_{t} \mid y_{t-1},\dots,y_{t-q}),  
\ese
where $p_{0}$ denotes the distribution of the initial $q$ variables $\by_{1:q}$; 
we follow common convention and condition on the initial observations to avoid modeling $p_0$.

We call such a sequence a Markov chain of maximal order $q$ if conditional on the values of $(y_{t-1},\dots,y_{t-q})$, the distribution of $y_{t}$ is independent of its more distant past, 
but the actual lags important in determining the distribution of $y_{t}$ may be an arbitrary subset of $(y_{t-1},\ldots,y_{t-q})$. In contrast, if the distribution of $y_{t}$ actually varies with the values at all the previous $q$ times points, we call the sequence a Markov chain of full order $q$. 
The case $q=0$ corresponds to serial independence. 

For a chain with $C_{0}$ states, there are $C_{0}-1$ free parameters in the conditional distribution of $y_{t}$, which can potentially vary arbitrarily with every possible combination of the levels of the previous variables. For a Markov chain of maximal order $q$, there are a total of $C_{0}^{q}$ such combinations, and hence the number of parameters in the full model is $(C_{0}-1)C_{0}^{q}$. This number increases exponentially in the order of the chain, creating estimation problems as $q$ increases.  It is very important to define flexible, parsimonious and interpretable representations, with unnecessary lags eliminated.  

A common approach to modeling higher order Markov chains is based on multinomial logit or probit models, with the lags included as linear predictors \citep{liang_zeger:1986, zeger_liang:1986}.
Modeling $r\th$ order interactions among the lags using such models would require the inclusion of ${q\choose r}(C_{0}-1)^{r}$ interaction terms in the set of linear predictors. The number of interaction terms thus increases rapidly with $C_{0}$ and $q$. 
For example, with only $5$ lags and $4$ categories, accommodation of second order interactions requires the inclusion of $90$ interaction terms. 
In practical applications, attention is thus often restricted to only a small number of lags and low order interaction terms \citep{fahrmeir_kaufmann:1987}. 

An alternative that can accommodate a relatively large number of lags but ignores interactions among lags is mixtures of transition distributions (MTD).  
In the basic MTD model \citep{raftery:1985}, the transition probability $p(y_{t} \mid y_{t-1},\dots,y_{t-q}) $ is a linear combination of 
$Q(y_{t-1},y_{t}),\dots,Q(y_{t-q},y_{t})$, 
where $Q$ is a $C_{0}\times C_{0}$ transition matrix for a first order Markov chain. 
\cite{raftery:1985b} and \citet{berchtold:1995, berchtold:1996} allowed different transition matrices for different lags.  
\cite{raftery_tavare:1994} and \cite{berchtold_raftery:2002} discussed estimation algorithms and other generalizations. 
While MTD leads to parsimonious models for higher order Markov chains, 
it is not structurally rich, particularly when size of the state space and/or the order of the chain is large. 
Additionally, the model implicitly assumes the process is of full order $q$, with selection of $q$ requiring refitting for different choices.

Another popular strategy to modeling higher order Markov chains is based on trees with conditioning sequences of different lengths as nodes and leaves. 
Variable length Markov chains (VLMC) \citep{buhlmann_wyner:1999, ron_etal:1996} prune large branches,  keeping only those nodes whose effects on $y_{t}$ are different enough from their parent's.  Context tree weighting \citep{willems_etal:1995} uses an ensemble of trees of varying depths. 
The sequence memoizer \citep{teh:2006, wood_etal:2011} uses a hierarchical prior to center the children $p(y_{t} \mid y_{t-1},\dots,y_{t-r})$ around their parent $p(y_{t} \mid y_{t-1},\dots,y_{t-(r-1)})$ for each $r\geq 1$, which favors a restrictive structure.  In general, tree based methods are not suitable when a more distant lag may be a more important predictor of $y_{t}$ than a relatively recent one. Sparse Markov chains (SMC) \citep{jaaskinen_etal:2014}  
attempt to remove this limitation.  SMCs cluster the lag combinations having similar influence on the transition distribution of $y_{t}$, related to VLMC but leaving the partitioning unrestricted.  Such hard clustering may lead to oversimplification of the dependence structure for long sequences.  Additionally, hard clustering and tree based approaches do not explicitly characterize significance of individual lags or provide a framework for testing of related hypotheses. 

In this article, we take a fundamentally different approach. Tensor factorizations for categorical regression have been developed in \cite{yang_dunson:2015}. 
We adapt these factorizations to our dynamic setting, while incorporating substantial improvements to the structure and computation.  
The proposed formulation leads to parsimonious representations of transition probability tensors, shrinking towards low dimensional structures and borrowing strength across lags, while being flexible in capturing complex higher order interactions.  
The method allows automated order and lag selection, quantifying uncertainty in selection and facilitating testing of hypotheses. 
Convergence of the posterior to the true transition probability tensor is guaranteed under ergodicity of the true data generating process. 
Taking a novel approach to posterior computation in variable dimension models, we develop an efficient Markov chain Monte Carlo (MCMC) algorithm.  

The article is organized as follows. 
Section \ref{sec: models} details our model and its interesting aspects. 
Section \ref{sec: estimation and inference} describes MCMC algorithms to sample from the posterior.
Section \ref{sec: simulation experiments} presents the results of simulation experiments comparing our method with existing approaches.
Section \ref{sec: applications} presents some applications of the proposed method. 
Section \ref{sec: discussion} contains concluding remarks.

\section{Model Specification}  \label{sec: models}
\subsection{Review of Tensor Factorizations} \label{sec: TF review}

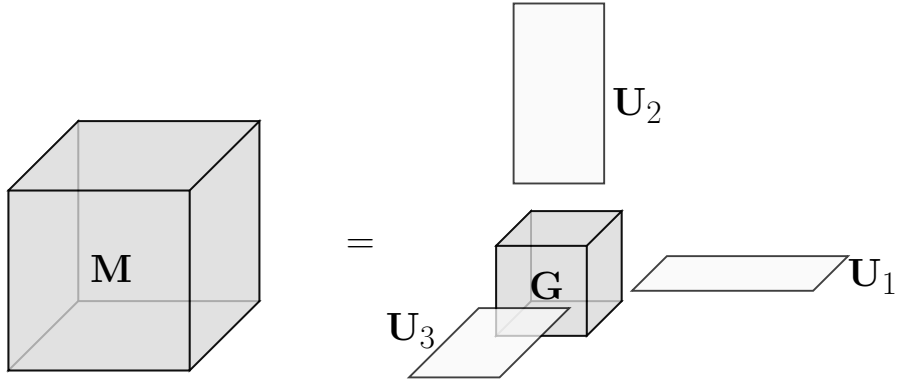
\begin{figure}[h!]
\centering
\begin{center}
\resizebox{12cm}{5cm}{%
\begin{tikzpicture}
	[
		grid/.style={very thin,gray},
		axis/.style={->,blue,thick},
		cube/.style={opacity=.75,very thick,fill=gray!25}]


	\draw[cube] (0,0,0) -- (0,4,0) -- (4,4,0) -- (4,0,0) -- cycle;
	
	\draw[cube] (0,0,0) -- (0,4,0) -- (0,4,4) -- (0,0,4) -- cycle;

	\draw[cube] (0,0,0) -- (4,0,0) -- (4,0,4) -- (0,0,4) -- cycle;

	\draw[cube] (4,0,0) -- (4,4,0) -- (4,4,4) -- (4,0,4) -- cycle;

	\draw[cube] (0,0,4) -- (0,4,4) -- (4,4,4) -- (4,0,4) -- cycle;

	\draw[cube] (0,4,0) -- (4,4,0) -- (4,4,4) -- (0,4,4) -- cycle;

\node[] at (7,2,2) {\Huge$=$};

	\draw[cube] (10,0,0) -- (10,0,2) -- (12,0,2) -- (12,0,0) -- cycle;
	
	\draw[cube] (10,0,0) -- (12,0,0) -- (12,2,0) -- (10,2,0) -- cycle;

	\draw[cube] (10,0,0) -- (10,0,2) -- (10,2,2) -- (10,2,0) -- cycle;

	\draw[cube] (12,0,0) -- (12,0,2) -- (12,2,2) -- (12,2,0) -- cycle;

	\draw[cube] (10,2,2) -- (10,0,2) -- (12,0,2) -- (12,2,2) -- cycle;

	\draw[cube] (10,2,0) -- (10,2,2) -- (12,2,2) -- (12,2,0) -- cycle;

	\draw[cube,fill=gray!5] (13,1,0) -- (13,1,2) -- (17,1,2) -- (17,1,0) -- cycle;		

	\draw[cube,fill=gray!5] (10,3,1) -- (12,3,1) -- (12,7,1) -- (10,7,1) -- cycle;	
	
	\draw[cube,fill=gray!5] (10,1,3) -- (10,1,7) -- (12,1,7) -- (12,1,3) -- cycle;

\node[] at (1.5,1.5,2) {\Huge$\bM$};
\node[] at (11.5,1.5,3) {\Huge$\bG$};
\node[] at (18.7,1.7,3) {\Huge$\bU_{1}$};
\node[] at (13.5,5.5,3) {\Huge$\bU_{2}$};
\node[] at (8.5,0.5,3) {\Huge$\bU_{3}$};

\end{tikzpicture}
}
\end{center}
\caption{Pictorial representation of HOSVD for a 3 way tensor $\bM$ with core tensor $\bG$ and weight matrices $\bU_{j}, j=1,2,3$.}
\label{fig: HOSVD}
\end{figure}

There is a vast literature on tensor factorizations, 
the two most popular approaches being parallel factor analysis (PARAFAC) and higher order singular value decomposition (HOSVD). 
PARAFAC \citep{harshman:1970} decomposes a $D_{1} \times \dots \times D_{p}$ dimensional tensor $\bM=\{m_{x_{1},\dots,x_{p}}\}$ as the sum of rank one tensors as
\be
m_{x_{1},\dots,x_{p}} = \sum_{h=1}^{k}g_{h} \prod_{j=1}^{p} u_{h}^{(j)}(x_{j}). 
\ee 
In contrast, HOSVD, proposed by \cite{tucker:1966} for three way tensors and extended to the general case by \cite{de_lathauwer_etal:2000}, 
factorizes $\bM$ as 
\be
m_{x_{1},\dots,x_{p}} = \sum_{h_{1}=1}^{k_{1}}\cdots\sum_{h_{p}=1}^{k_{p}}g_{h_{1},\dots,h_{p}} \prod_{j=1}^{p} u_{h_{j}}^{(j)}(x_{j}), 
\ee
where $\bG=\{g_{h_{1},\dots,h_{p}}\}$, called a core tensor, captures interactions between the different components and $\bU_{j}=\{u_{h_{j}}^{(j)}(x_{j})\}$ are component specific weights.  See Figure \ref{fig: HOSVD}.  HOSVD achieves better data compression and requires fewer components compared to PARAFAC, which can be obtained as a special case of HOSVD with $\bG$ diagonal.

The tensor factorization that is most relevant to our problem was introduced in \cite{yang_dunson:2015} (YD). 
YD considered the problem of regressing a categorical response variable $y \in \{1,\dots,D_{0}\}$ on categorical predictors $x_{j} \in \{1,\dots,D_{j}\}$, $j=1,\dots,p$. 
Structuring the conditional probabilities $p(y\mid x_{j},j=1,\dots,p)$ as the elements of a $D_{0} \times D_{1} \times \dots \times D_{p}$ dimensional tensor, 
YD proposed the following HOSVD-type factorization 
\be
p(y\mid x_{j},j=1,\dots,p)   =   \sum_{h_{1}=1}^{k_{1}}\cdots\sum_{h_{p}=1}^{k_{p}} \lambda_{h_{1}\dots h_{p}}(y)\prod_{j=1}^{p}\pi_{h_{j}}^{(j)}(x_{j}),  \label{eq: YD CTF}
\ee
where $1 \leq k_{j} \leq D_{j}$ for $j=1,\dots,p$ and the parameters $\lambda_{h_{1},\dots,h_{p}}(y)$ and $\pi_{h_{j}}^{(j)}(x_{j})$ are all non-negative and satisfy the constraints 
(a) $\sum_{y=1}^{D_{0}}  \lambda_{h_{1}\dots h_{p}}(y) =1$ for each combination $(h_{1},\dots,h_{p})$, and 
(b) $\sum_{h_{j}=1}^{k_{j}} \pi_{h_{j}}^{(j)}(x_{j}) = 1$ for each pair $(j,x_{j})$.
They established that any conditional probability tensor can be represented as (\ref{eq: YD CTF}), with the parameters satisfying the constraints (a) and (b). 
The constraints (a) and (b) are thus not restrictive but they ensure that $\sum_{y=1}^{D_{0}} p(y\mid x_{j},j=1,\dots,p)=1$.

Taking a Bayesian approach, 
they assigned sparsity inducing priors on the $k_{j}$'s and conditional on the $k_{j}$'s, placed independent Dirichlet priors on $\lambda_{h_{1},\dots,h_{p}}(y)$'s and $\pi_{h_{j}}^{(j)}(x_{j})$'s as
$\{\lambda_{h_{1},\dots,h_{q}}(1),\dots,\lambda_{h_{1},\dots,h_{p}}(D_{0})\} \sim \Dir(1/D_{0},\dots,1/D_{0})$ for each combination $(h_{1},\dots,h_{p})$ with $1 \leq h_{j} \leq k_{j}$ 
and $\{\pi_{1}^{(j)}(x_{j}),\dots,\pi_{k_{j}}^{(j)}(x_{j})\} \sim \Dir(1/k_{j},\dots,1/k_{j})$ for each $x_{j} \in \{1,\dots,D_{j}\}$.
The dimensions of these parameters vary with $k_{j}$'s, making the design of efficient MCMC algorithms  challenging. YD used an approximate two-stage sampler, selecting the $k_{j}$'s in the first stage and then sampling the other parameters in the second stage while keeping the $k_{j}$'s fixed. 

\subsection{Higher Order Markov Chains via Tensor Factorization}  \label{sec: HOMC via CTF}
We propose a nonparametric Bayes approach for inferring the order and structure of higher order Markov chains building on a YD-type conditional tensor factorization.  In our dynamic setting, we have a time-indexed categorical sequence $\{y_{t}\}$ with finite memory of maximal order $q$ taking values in the set $\{1,\dots,C_{0}\}$. 
Given $y_{t-1},\ldots,y_{t-q}$, the distribution of $y_{t}$ is independent of all observations prior to $t-q$. 
The variables that are important in predicting $y_t$ can potentially constitute a subset of $\{y_{t-1},\ldots,y_{t-q}\}$. 
For $t=q+1,\dots,T$, the transition probability $p(y_{t} \mid y_{t-1},\ldots,y_{t-q})$ is structured as a $C_{0} \times C_{0} \times \dots \times C_{0}$ dimensional tensor and admits the factorization   
\be
&&\hspace{-1cm} p(y_{t} \mid y_{t-j},j=1,\dots,q)  = \sum_{h_{1}=1}^{k_{1}}\cdots\sum_{h_{q}=1}^{k_{q}} \lambda_{h_{1},\dots,h_{q}}(y_{t})\prod_{j=1}^{q}\pi_{h_{j}}^{(j)}(y_{t-j}), \label{eq: TFM1}
\ee
where, with some repetition, $1 \leq k_{j} \leq C_{0}$ for all $j$ and the parameters $\lambda_{h_{1},\dots,h_{p}}(y_{t})$ and $\pi_{h_{j}}^{(j)}(y_{t-j})$ are all non-negative and satisfy the constraints 
\be
&&\sum_{y_{t}=1}^{C_{0}}  \lambda_{h_{1},\dots,h_{q}}(y_{t}) =1,~~ \text{for each combination}~(h_{1},\dots,h_{q}),  \label{eq: TFM2}\\
&&\sum_{h_{j}=1}^{k_{j}} \pi_{h_{j}}^{(j)}(y_{t-j}) = 1, ~~ \text{for each pair }~(j,y_{t-j}). \label{eq: TFM3}
\ee

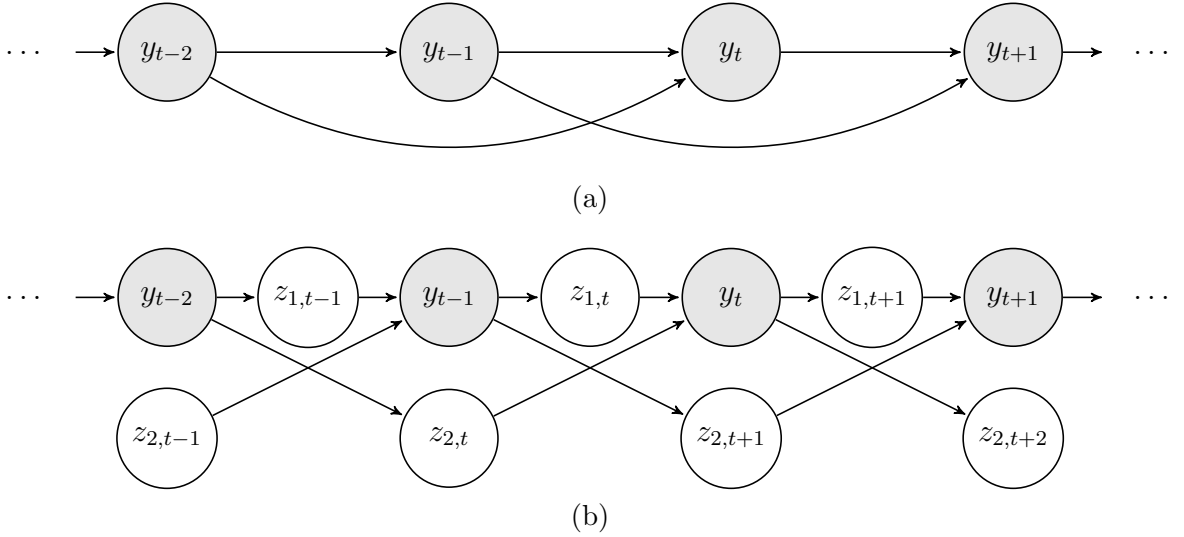
\begin{figure}[h!]
\subfloat[]
{
\centering
\begin{tikzpicture}[scale=1.5,->,>=stealth',shorten >=1pt,auto,node distance=2.8cm,semithick]

  \node[minimum size=1.3cm] (y_t-3) at (-0.75,0) {$\dots$};
  \node[style={draw,circle,fill=gray!20},minimum size=1.3cm] (y_t-2) at (0.5,0) {$y_{t-2}$};
  \node[style={draw,circle,fill=gray!20}, minimum size=1.3cm] (y_t-1) at (3,0) {$y_{t-1}$};
  \node[style={draw,circle,fill=gray!20}, minimum size=1.3cm] (y_t) at (5.5,0) {$y_{t}$};
  \node[style={draw,circle,fill=gray!20}, minimum size=1.3cm] (y_t+1) at (8,0) {$y_{t+1}$};
  \node[minimum size=1.3cm] (y_t+2) at (9.25,0) {$\dots$};

  \path (y_t-3) edge (y_t-2);
  \path (y_t-2) edge (y_t-1);
  \path (y_t-1) edge (y_t);
  \path (y_t) edge (y_t+1);
  \path (y_t+1) edge (y_t+2);

  \path (y_t-2) edge[bend right] (y_t);
  \path (y_t-1) edge[bend right] (y_t+1);

\end{tikzpicture}
}

\subfloat[]
{
\begin{tikzpicture}[scale=1.5,->,>=stealth',shorten >=1pt,auto,node distance=2.8cm,semithick]

  \node[style={draw,circle,fill=gray!20},minimum size=1.3cm] (y_t-2) at (0.5,0) {$y_{t-2}$};
  \node[style={draw,circle,fill=gray!20}, minimum size=1.3cm] (y_t-1) at (3,0) {$y_{t-1}$};
  \node[style={draw,circle,fill=gray!20}, minimum size=1.3cm] (y_t) at (5.5,0) {$y_{t}$};
  \node[style={draw,circle,fill=gray!20}, minimum size=1.3cm] (y_t+1) at (8,0) {$y_{t+1}$};

  \node[minimum size=1.3cm] (z_1_t-2) at (-0.75,0) {$\dots$};
  \node[style={draw,circle}, minimum size=1.3cm] (z_2_t-1) at (0.5,-1.25) {$z_{2,t-1}$};
  \node[style={draw,circle}, minimum size=1.3cm] (z_1_t-1) at (1.75,0) {$z_{1,t-1}$};
  \node[style={draw,circle}, minimum size=1.3cm] (z_2_t) at (3,-1.25) {$z_{2,t}$};
  \node[style={draw,circle}, minimum size=1.3cm] (z_1_t) at (4.25,0) {$z_{1,t}$};
  \node[style={draw,circle}, minimum size=1.3cm] (z_2_t+1) at (5.5,-1.25) {$z_{2,t+1}$};
  \node[style={draw,circle}, minimum size=1.3cm] (z_1_t+1) at (6.75,0) {$z_{1,t+1}$};
  \node[style={draw,circle}, minimum size=1.3cm] (z_2_t+2) at (8,-1.25) {$z_{2,t+2}$};
  \node[minimum size=1.3cm] (z_1_t+2) at (9.25,0) {$\dots$};

 \path (z_1_t-2) edge (y_t-2);
 \path (z_2_t-1) edge (y_t-1);
 \path (z_1_t-1) edge (y_t-1);
  \path (z_2_t) edge (y_t);
  \path (z_1_t) edge (y_t);
  \path (z_2_t+1) edge (y_t+1);
  \path (z_1_t+1) edge (y_t+1);

  \path (y_t-2) edge (z_1_t-1);
  \path (y_t-1) edge (z_1_t);
  \path (y_t) edge (z_1_t+1);
  \path (y_t+1) edge (z_1_t+2);
  
  \path (y_t-2) edge (z_2_t);
  \path (y_t-1) edge (z_2_t+1);
  \path (y_t) edge (z_2_t+2);

\end{tikzpicture}
}
\caption{Graphical model depicting the dependence structure of a second order Markov chain $\{y_{t}\}$ (a) without and (b) with latent variables $\{z_{j,t}\}$.}
\label{fig: graph 2}
\end{figure}

Introducing latent allocation variables $z_{j,t}$ for each $j=1,\dots,q$ and $t=q+1,\dots,T$, the response values are conditionally independent 
and the factorization can be equivalently represented through the following hierarchical formulation:
\be
(y_{t}\mid z_{j,t}=h_{j},j=1,\dots,q)   	&\sim&   \Mult(\{1,\dots,C_{0}\},\lambda_{h_{1},\dots,h_{q}}(1),\dots,\lambda_{h_{1},\dots,h_{q}}(C_{0})), \label{eq: interaction}    \\
(z_{j,t}\mid y_{t-j})   	&\sim&   \Mult(\{1,\dots,k_{j}\},\pi_{1}^{(j)}(y_{t-j}),\dots \pi_{k_{j}}^{(j)}(y_{t-j})).  \label{eq: soft clustering}
\ee
See Figure \ref{fig: graph 2}.  Posterior computation is facilitated by sampling these latent auxiliary variables. 
This formulation also aids in understanding interesting features of the model. 
Equation (\ref{eq: soft clustering}), for instance, reveals the soft clustering property of the model that enables it to borrow strength across the different categories of $y_{t-j}$ 
by allowing the $z_{j,t}$'s associated with a particular state of $y_{t-j}$ to be allocated 
to different latent populations, which are shared across all $C_{0}$ states of $y_{t-j}$. 
Equation (\ref{eq: interaction}), on the other hand, shows how such soft assignment enables the model to capture complex interactions among the lags in an implicit and parsimonious manner by allowing the latent populations indexed by $(h_{1},\dots,h_{q})$ to be shared among the various state combinations of the lags.  

When $k_{j}=1$, $\pi_{1}^{(j)}(y_{t-j})=1$ and $p(y_{t} \mid y_{t-1},\ldots,y_{t-q})$ does not vary with $y_{t-j}$. 
The variable $k_{j}$ thus determines the inclusion of the $j\th$ lag $y_{t-j}$ in the model. 
The variable $k_{j}$ also determines the number of latent classes for the $j\th$ lag $y_{t-j}$. 
The number of parameters in such a factorization is given by $(C_{0}-1)\prod_{j=1}^{q}k_{j} + C_{0}\sum_{j=1}^{q}(k_{j}-1)$, which will be much smaller than the number of parameters $(C_{0}-1)C_{0}^{q}$ required to specify a full Markov model of the same maximal order if $\prod_{j=1}^{q}k_{j} \ll C_{0}^{q}$.

In practical applications $\prod_{j=1}^{q}k_{j}$ may still be quite large. 
For instance, for a Markov chain with $C_{0}=4$ states and $5$ important lags with $k_{j}=3$ for all $j=1,\dots,5$, 
the number of parameters required to specify the core tensor will be $3 \times 3^{5}=729$. 
While this results in a significant reduction in the number of parameters compared to a fully specified Markov model of the same maximal order which requires $3 \times 4^5 = 3072$ parameters, 
it may still be too large for efficient and numerically stable estimation of the parameters for data sets of sizes that are typically encountered in practice.  

Towards a more parsimonious representation,  we note that the conditional tensor factorization (\ref{eq: TFM1}) can be interpreted as a predictor dependent mixture model for modeling distributions supported on $\{1,\dots,C_{0}\}$.  
Here the probability vectors $\blambda_{h_{1},\dots,h_{q}}=\{\lambda_{h_{1},\dots,h_{q}}(1),\dots, \lambda_{h_{1},\dots,h_{p}}(C_{0})\}$ that constitute the core tensor play the role of kernels of the mixture model, 
and $\pi_{h_{1},\dots,h_{q}}(y_{t-1},\dots,y_{t-q})=\prod_{j=1}^{q}\pi_{h_{j}}^{(j)}(y_{t-j})$ play the role of associated predictor dependent mixture weights. 
Given $k_{1},\dots,k_{q}$, the kernels are indexed by $(h_{1},\dots,h_{q})$ with $h_{j}=1,\dots,k_{j}$, contributing $\prod_{j=1}^{q}k_{j}$ mixture components to the model. 
Thus, the number of kernels determines the effective dimension of the model. 
In most applications, a very large number of kernels may not be required. 
This is especially true for discrete distributions supported on a finite set $\{1,\dots,C_{0}\}$.  

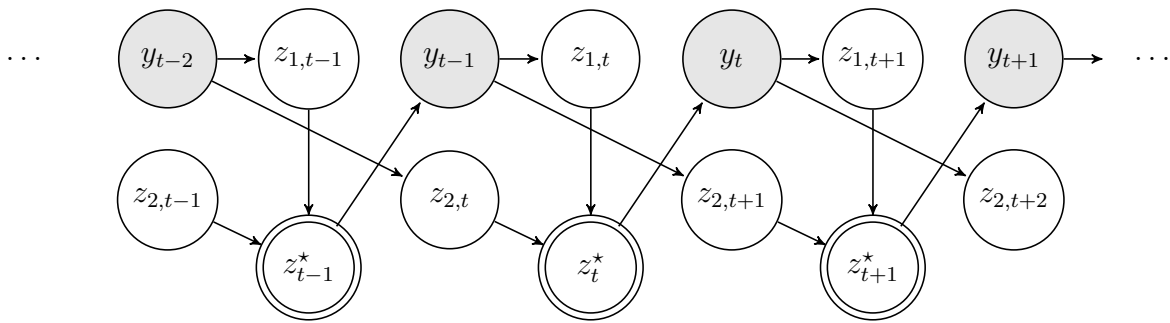
\begin{figure}[ht!]
\begin{tikzpicture}[scale=1.5,->,>=stealth',shorten >=1pt,auto,node distance=2.8cm,semithick]

  \node[style={draw,circle,fill=gray!20},minimum size=1.3cm] (y_t-2) at (0.5,0) {$y_{t-2}$};
  \node[style={draw,circle,fill=gray!20}, minimum size=1.3cm] (y_t-1) at (3,0) {$y_{t-1}$};
  \node[style={draw,circle,fill=gray!20}, minimum size=1.3cm] (y_t) at (5.5,0) {$y_{t}$};
  \node[style={draw,circle,fill=gray!20}, minimum size=1.3cm] (y_t+1) at (8,0) {$y_{t+1}$};

  \node[minimum size=1.3cm] (z_1_t-2) at (-0.75,0) {$\dots$};
  \node[style={draw,circle}, minimum size=1.3cm] (z_2_t-1) at (0.5,-1.25) {$z_{2,t-1}$};
  \node[style={draw,circle}, minimum size=1.3cm] (z_1_t-1) at (1.75,0) {$z_{1,t-1}$};
  \node[style={draw,circle}, minimum size=1.3cm] (z_2_t) at (3,-1.25) {$z_{2,t}$};
  \node[style={draw,circle}, minimum size=1.3cm] (z_1_t) at (4.25,0) {$z_{1,t}$};
  \node[style={draw,circle}, minimum size=1.3cm] (z_2_t+1) at (5.5,-1.25) {$z_{2,t+1}$};
  \node[style={draw,circle}, minimum size=1.3cm] (z_1_t+1) at (6.75,0) {$z_{1,t+1}$};
  \node[style={draw,circle}, minimum size=1.3cm] (z_2_t+2) at (8,-1.25) {$z_{2,t+2}$};
  \node[minimum size=1.3cm] (z_1_t+2) at (9.25,0) {$\dots$};

  \node[style={double,double distance=2pt,draw,circle}, minimum size=1.3cm] (z_t-1_star) at (1.75,-1.85) {$z_{t-1}^{\star}$};
  \node[style={double,double distance=2pt,draw,circle}, minimum size=1.3cm] (z_t_star) at (4.25,-1.85) {$z_{t}^{\star}$};
  \node[style={double,double distance=2pt,draw,circle}, minimum size=1.3cm] (z_t+1_star) at (6.75,-1.85) {$z_{t+1}^{\star}$};

 \path (z_2_t-1) edge (z_t-1_star);
 \path (z_1_t-1) edge (z_t-1_star);
  \path (z_2_t) edge (z_t_star);
  \path (z_1_t) edge (z_t_star);
  \path (z_2_t+1) edge (z_t+1_star);
  \path (z_1_t+1) edge (z_t+1_star);

  \path (y_t-2) edge (z_1_t-1);
  \path (y_t-1) edge (z_1_t);
  \path (y_t) edge (z_1_t+1);
  \path (y_t+1) edge (z_1_t+2);
  
  \path (y_t-2) edge (z_2_t);
  \path (y_t-1) edge (z_2_t+1);
  \path (y_t) edge (z_2_t+2);

 \path (z_t-1_star) edge (y_t-1);
  \path (z_t_star) edge (y_t);
  \path (z_t+1_star) edge (y_t+1);

\end{tikzpicture}
\caption{Graphical model depicting the dependence structure of a second order Markov chain $\{y_{t}\}$ with latent variables $\{z_{j,t}\}$ and $\{z_{t}^{\star}\}$.}
\label{fig: graph 3}
\end{figure}

A more parsimonious representation that retains the flexibility of the original model is obtained by encouraging the kernels $\blambda_{h_{1},\dots,h_{q}}$ to be shared amongst the label combinations $(h_{1},\dots,h_{q})$ through probabilistic clustering. 
Specifically, we let 
\be
\hspace{-1cm} \blambda_{h_{1},\dots,h_{q}}    &\sim&   \sum_{\ell=1}^{\infty} \pi_{\ell}^{\star}\delta_{\blambda_{\ell}^{\star}},~~\hbox{independently for each}~(h_{1},\dots,h_{q}),  \label{eq: hDP1} \\
\hspace{-1cm}\blambda_{\ell}^{\star} = \{\lambda_{\ell}^{\star}(1),\dots,\lambda_{\ell}^{\star}(C_{0})\}    &\sim&   \Dir(\alpha,\dots,\alpha),~~\hbox{independently for}~\ell=1,\dots,\infty,  ~ \label{eq: hDP2}\\
\hspace{-1cm}\pi_{\ell}^{\star}  = V_{\ell}\prod_{m=1}^{\ell-1}(1-V_{m}), ~~~~~ V_{\ell} &\sim& \Beta(1,\alpha_{0}),~~\hbox{independently for}~\ell=1,\dots,\infty.  \label{eq: hDP3}
\ee
Introducing latent variables $z_{t}^{\star}$ and $z_{h_{1},\dots,h_{q}}^{\star}$, 
for $t=t^{\star},\dots,T$, $h_{j}=1,\dots,k_{j}$, $j=1,\dots,q$, 
we further have
\be
&& p(z_{h_{1},\dots,h_{q}}^{\star}=\ell) ~=~ \pi_{\ell}^{\star},~~\hbox{independently for each}~(h_{1},\dots,h_{q}),\\
&& (\blambda_{h_{1},\dots,h_{q}} \mid z_{h_{1},\dots,h_{q}}^{\star}=\ell) ~=~ \blambda_{\ell}^{\star},  \label{eq: hDP2 latent variables} ~~~~~~(z_{t}^{\star}\mid z_{j,t}=h_{j},j=1,\dots,q)=z_{h_{1},\dots,h_{q}}^{\star}, \\  
&& (y_{t} \mid z_{t}^{\star}=\ell)   	~\sim~   \Mult(\{1,\dots,C_{0}\},\lambda_{\ell}^{\star}(1),\dots,\lambda_{\ell}^{\star}(C_{0})). 
\ee
See Figure \ref{fig: graph 3}. 
The cluster inducing prior specified through (\ref{eq: hDP1})-(\ref{eq: hDP3}) corresponds to a Dirichlet process (DP) prior \citep{ferguson:1973} written in terms of its stick-breaking representation \citep{sethuraman:1994}.   
Although the prior allows infinitely many components, the number of components occupied by the $\prod_{j=1}^{q}k_{j}$ mixture kernels is finite and likely much smaller than $\prod_{j=1}^{q}k_{j}$, leading to a significant reduction in the effective number of parameters of the model.
Our experiments suggest that, even in low to moderate dimensional problems, such clustering of kernels greatly improves numerical stability compared with assigning continuous priors on the kernels.  The idea of hierarchical sharing of the kernels constituting the core tensor is not specific to our dynamic setting, and can be easily adapted to other tensor factorization models including the original YD model, also eliminating problems with exceeding limited storage space that plague YD in applications we have considered.

The DP prior on the mixture kernels $\blambda_{h_{1},\dots,h_{q}}$ treats the kernel indices $(h_{1},\dots,h_{q})$ as exchangeable.  Although it is conceptually appealing to favor clustering of kernels with similar indices $(h_{1},\dots,h_{q})$, there is an associated significant increase in model complexity and computational costs.  Hence, given that exchangeable priors worked well in examples we considered, we did not consider such modifications further.  Although we focused on a DP prior for simplicity, other exchangeable clustering priors, such as Pitman-Yor processes \citep{pitman_yor:1997} or probit stick-breaking processes \citep{rodriguez_dunson:2011}, can be used.

Next, consider mixture probability vectors $\bpi_{k_{j}}^{(j)}(y_{t-j})=\{\pi_{1}^{(j)}(y_{t-j}),\dots,\pi_{k_{j}}^{(j)}(y_{t-j})\}$. 
The dimension of $\bpi_{k_{j}}^{(j)}(y_{t-j})$, unlike the $\blambda_{h_{1},\dots,h_{q}}$'s, varies linearly with $k_{j}$. 
For a Markov chain with $y_{0}=4$ states and $5$ important lags with $k_{j}=3$ for all $j=1,\dots,5$, 
the number of parameters contributed by all the $\bpi_{k_{j}}^{(j)}(y_{t-j})$ vectors will only be $4 \times 2 \times 5=40$. 
We thus assign independent priors on $\bpi_{k_{j}}(y_{t-j})$ as 
\be
\bpi_{k_{j}}^{(j)}(y_{t-j}) = \{\pi_{1}^{(j)}(y_{t-j}),\dots,\pi_{k_{j}}^{(j)}(y_{t-j})\} &\sim& \Dir(\gamma_{j},\dots,\gamma_{j}).  \label{eq: prior on pi}
\ee
The probability vectors $\bpi_{k_{j}}^{(j)}(y_{t-j})$ are supported on $\{1,\dots,k_{j}\}$ for each pair $(j,y_{t-j})$. 
Therefore, unlike for $\blambda_{h_{1},\dots,h_{q}}$, conditioning on $k_{j}$, which we have kept implicit in (\ref{eq: prior on pi}), can not be avoided. 
We, however, do not allow the hyper-parameter $\gamma_{j}$ to vary with $k_{j}$. 
This has important consequences in the design of our MCMC sampler. 
Details are deferred to Section \ref{sec: posterior computation}.

Finally, model specification is completed by assigning priors on $\bk$. 
We assign independent priors on $k_{j}$'s as 
\be
p(k_{j}=k) &=& p_{0,j}(k) ~ \propto ~ \exp(-\varphi jk),~~~j=1,\dots,q,~k=1,\dots,C_{0},     \label{eq: prior on k_{j}'s}
\ee
where $\varphi>0$.  
The prior assigns increasing probabilities to smaller values of $k_{j}$ as the lag $j$ becomes more distant, 
reflecting the natural belief that increasing lags will have diminishing influence on the distribution of $y_{t}$. 
Larger values of $\varphi$ imply faster decay of $p_{0,j}(k)$ with increase in $j$ and $k$, favoring sparser models. See Figure \ref{fig: varphi} in the Supplementary Materials. 
The model space can be restricted to the class of Markov models of full order by restricting the prior to satisfy the condition that $k_{j+1}=1$ whenever $k_{j}=1$ for some $j$. It is appealing to avoid such restrictions to accommodate scenarios 
where a more distant lag is an important predictor of $y_{t}$ but a lag in the more recent past is not.   
As illustrated in Section \ref{sec: applications}, such scenarios are often encountered in practice.   

Let $\by=\{y_{t}: t=t^{\star},\dots,T\}$, $\bz=\{z_{j,t}: t=t^{\star},\dots,T,j=1,\dots,q\}$ 
and $\bz^{\star}=\{z_{h_{1},\dots,h_{q}}^{\star}: h_{j}=1,\dots,k_{j},j=1,\dots,q\}$, 
where $t^{\star}=(q+1)$. 
Collecting all potential predictors of $y_{t}$ in $\bw_{t}=(w_{1,t},\ldots,w_{q,t})\trans$ with $w_{j,t} = y_{t-j}$ for $j=1,\ldots,q$ and $t=t^{\star},\dots,T$, 
the joint distribution of $\by$, $\bz$ and $\bz^{\star}$ admits the factorization
\be
&&\hspace{-1cm} p(\by,\bz,\bz^{\star} \mid \blambda^{\star},\bpi^{\star},\bpi_{\bk},\bk) = \prod_{t=t^{\star}}^{T} \left\{p(y_{t} \mid \blambda_{z_{\bz_{t}}^{\star}}^{\star}) \prod_{j=1}^{q} p(z_{j,t} \mid w_{j,t},\bpi_{k_{j}}^{(j)},k_{j}) \right\} \prod_{j=1}^{q}\prod_{h_{j}=1}^{k_{j}}p(z_{h_{1},\dots,h_{q}}^{\star} \mid \bpi^{\star})     \nonumber\\
&& = \prod_{t=t^{\star}}^{T} \{p(y_{t} \mid \blambda_{z_{\bz_{t}}^{\star}}^{\star}) ~ p(\bz_{t} \mid \bw_{t},\bpi_{\bk},\bk) \}   ~  p(\bz^{\star} \mid \bpi^{\star})   \nonumber\\
&& = p(\by \mid \bz,\bz^{\star},\blambda^{\star}) ~ \prod_{j=1}^{q}p(\bz_{j} \mid \bw_{j}, \bpi_{k_{j}}^{(j)}, k_{j})  ~  p(\bz^{\star} \mid \bpi^{\star}).  \label{eq: p(Y,Z,Zstar) factorization}
\ee
%
%
%
Here $\bk=\{k_{j}: j=1,\dots,q\}$, 
$\bpi_{k_{j}}^{(j)}(w_{j})=\{\pi_{h_{j}}^{(j)}(w_{j}): h_{j}=1,\dots,k_{j}\}$, 
$\bpi_{k_{j}}^{(j)}=\{\bpi_{k_{j}}^{(j)}(w_{j}): w_{j}=1,\dots,C_{0}\}$, 
$\bpi_{\bk}=\{\bpi_{k_{j}}^{(j)}: j=1,\dots,q\}$. 
Also, $\bz_{t}=\{z_{j,t}: j=1,\dots,q\}$ for all $t=t^{\star},\dots,T$, $\bz_{j}=\{z_{j,t}: t=t^{\star},\dots,T\}$ for $j=1,\dots,q$ and $\bw_{j}=\{w_{j,t}: t=t^{\star},\dots,T\}$. 
Here $\blambda^{\star}=\{\blambda_{\ell}^{\star}: \ell=1,\dots,\infty\}$ and $\bpi^{\star}=\{\pi_{\ell}^{\star}: \ell=1,\dots,\infty\}$ collect respectively the atoms and the probabilities of distribution (\ref{eq: hDP1}). 
The suffixes $\bk$ and $k_{j}$ signify that the supports and hence the dimensions of the associated parameters depend on them.  

The proposed model is nonparametric in the sense that it assigns positive probability to neighborhoods of the true data generating process. 
Let $P_{0}$ denote the true transition probability tensor. 
Also let $d$ denote the $L_{1}$ distance between two transition probability tensors $P$ and $P_{0}$ defined as
\bse
d(P,P_{0}) = \sum_{w_{1}=1}^{C_{0}}\dots\sum_{w_{q}=1}^{C_{0}}\sum_{y=1}^{C_{0}}|P(y\mid w_{1},\dots, w_{q})-P_{0}(y \mid w_{1},\dots,w_{q})|.
\ese 
Let $\Pi$ denote the prior on the space of all transition probability tensors induced through the proposed model 
and $\Pi(\cdot \mid \by_{1:T})$ denote the corresponding posterior based on an observed sequence $\by_{1:T}$ of length $T$. 
The following result establishes posterior consistency of the proposed model under the mild assumption of ergodicity of the true data generating mechanism by showing that 
$\Pi(\cdot \mid \by_{1:T})$ concentrates in arbitrarily small $L_{1}$ neighborhoods of $P_{0}$ as the sequence length approaches infinity. 

\begin{Thm} \label{Thm: Main Theorem 0}
If the true data generating process is an ergodic Markov chain of maximal order $q$, 
then for any $\delta>0$, 
$\Pi\left\{P: \textstyle d(P,P_{0})>\delta \mid \by_{1:T}\right\} \to 0$ as $T\to \infty$ almost surely $P_{0}$. 
\end{Thm}

The proof, deferred to \ref{appendix: consistency}, follows along the lines of the proof of Theorem 4.3.1 of \cite{ghosh_ramamoorthy:2003}
and utilizes strong law of large numbers for ergodic Markov chains. 

We conclude this section by comparing the proposed approach to sparse Markov chains (SMC) \citep{jaaskinen_etal:2014}.  The SMC model can be formulated as 
\begin{eqnarray}
p(y_{t} \mid y_{t-j}, j=1,\dots,q) = \sum_{h=1}^{k}\lambda_{h}(y_{t}) \pi_{h}(y_{t-1},\ldots,y_{t-q}), \label{eq:SMC}
\end{eqnarray}
where $\sum_{y=1}^{C_{0}}\lambda_{h}(y)=1$ for $h=1,\dots,k$, $\pi_{h}(y_{t-1},\ldots,y_{t-q})=1$ if $(y_{t-1},\ldots,y_{t-q}) \in S_{h}$ and $0$ otherwise, 
and $\{ S_{h} \}_{h=1}^k$ forms an unrestricted partition of the set of all possible values of the conditioning sequence $(y_{t-1},\ldots,y_{t-q})$. 
Equation (\ref{eq:SMC}) is a predictor dependent mixture model induced via a  PARAFAC-type conditional tensor factorization.  
Introducing latent variables $z_{t}$, the model can be reformulated as
\bse
(y_{t} \mid z_{t}=h) &\sim& \Mult(\{1,\dots,C_{0}\},\lambda_{h}(1),\dots,\lambda_{h}(C_{0})), \\
(z_{t} \mid  y_{t-j}, j=1,\dots,q) &\sim& \Mult(\{1,\dots,k\},\pi_{1}(y_{t-1},\ldots,y_{t-q}),\dots,\pi_{k}(y_{t-1},\ldots,y_{t-q})).
\ese
By forcing $z_{t_{1}}=z_{t_{2}}$ whenever $(y_{t_{1}-1},\dots,y_{t_{1}-q})=(y_{t_{2}-1},\dots,y_{t_{2}-q})$, 
the model only allows a restrictive hard clustering of the mixture kernels. 
Additionally, the assumption of conditional independence of $y_t$ and $(y_{t-1},\ldots,y_{t-q})$ given a single latent variable $z_{t}$ is quite restrictive, and precludes inferences on the importance of individual lags.  Approaches making a similar assumption in the continuous time series literature, such as the 
model of \cite{di_lucca_etal:2013}, face similar disadvantages.

The model proposed in this article is based on a more general HOSVD-type conditional tensor factorization. 
The mixture components are indexed by a vector of indices $(h_{1},\dots,h_{q})$, not a single scalar index $h$, 
and the mixture probabilities admit a further decomposition as $\pi_{h_{1},\dots,h_{q}}(y_{t-1},\dots,y_{t-q})=\prod_{j=1}^{q}\pi_{h_{j}}^{(j)}(y_{t-j})$. 
The latent variable formulation, given by (\ref{eq: interaction})-(\ref{eq: soft clustering}), thus introduces a separate latent cluster indicator variable $z_{j,t}$ for each lag $y_{t-j}$.
This allows explicit characterization of the importance of individual lags through the variables $k_{j}$'s.
The variables $z_{j,t}$'s are allocated to different clusters in a soft probabilistic manner - 
for $t_{1} \neq t_{2}$, $z_{j,t_{1}}$ and $z_{j,t_{2}}$ are allowed to take different values even when $y_{t_{1}-j}=y_{t_{2}-j}$.
The cluster inducing DP prior on the mixture kernels $\blambda_{h_{1},\dots,h_{q}}$ provides opportunities for an additional layer of dimension reduction. 
These features enable the proposed model to capture complex serial dependence structures with greater parsimony, making it better suited to high-dimensional applications while also facilitating automated lag and order selection.

\section{Estimation and Inference} \label{sec: estimation and inference}
\subsection{Posterior Computation} \label{sec: posterior computation}
A mixture of finite mixture (MFM) model has a finite but unknown number of mixture components.  Our proposed model can be viewed as a sophisticated dynamic MFM model, with lag-specific number of mixture components $k_{j}$ and mixture probabilities $\bpi_{k_{j}}^{(j)}(w_{j})$. 
The dimension of $\bpi_{\bk}$ varies with $\bk$. 
The most common approach to posterior computation in variable-dimensional mixture models is reversible jump MCMC \citep{richardson_green:1997}.  Alternative algorithms include the allocation sampler \citep{nobile_fearnside:2007} and birth-death MCMC \citep{stephens:2000}.  It is difficult to design efficient implementations of such algorithms including in our setting.
To bypass this problem, \cite{yang_dunson:2015} developed an approximate two-stage sampler. 
In the first stage, a stochastic variable search algorithm \citep{george_mcculloch:1997} based on an approximated marginal likelihood was used 
to estimate the set of important predictors and corresponding values of $k_{j}$'s.  
In the second stage, samples of $\lambda_{h_{1},\dots,h_{p}}$ and $\pi_{h_{j}}^{(j)}(x_{j})$ were drawn from their closed form full conditionals conditionally on the estimated $k_j$'s.

Recently \cite{miller_harrison:2015} studied MFM models in the univariate iid case drawing parallels with infinite mixture models. 
Using a Dirichlet prior on the mixture weights $\bpi_{k}=(\pi_{1},\dots,\pi_{k}) \sim \Dir(\gamma,\dots,\gamma)$, they integrated out $\bpi_{k}$ and $k$ 
to obtain closed form expressions for the induced cluster configurations.  
Mimicking MCMC algorithms for Dirichlet process mixture (DPM) models, 
they developed a sampler that iterates between updating the cluster configurations conditional on the other parameters and then updating the other parameters conditional on the cluster configurations, bypassing the problem of defining proposals for the variable dimensional parameter $\bpi_{k}$.  The ability to marginalize is largely unique to simple Dirichlet mixture models, precluding a straightforward adaptation of their algorithm to our model. 
However, we were able to generalize their algorithm to not require marginalization by explicitly sampling $\bk$ from the posterior.
Given $\bk$, $\bpi_{\bk}$ is of fixed dimension and all parameters can be easily updated using standard techniques. 
The sampling of $\bk$ is thus a key innovative step in our sampler and we outline below how we do this. 
Technical details are deferred to \ref{appendix: posterior of k}.  

When the hyper-parameter $\gamma_{j}$ does not depend on $k_{j}$, 
it is possible to integrate out $\bpi_{k_{j}}^{(j)}$ to obtain closed form expressions for $p(k_{j} \mid \bz_{j}, \bw_{j})$.
Specifically, we have
\be
&&\hspace{-1cm} p(k_{j}\mid \bz_{j}, \bw_{j}) 
= \frac{  p_{0,j}(k_{j})\prod_{r=1}^{C_{0}}\frac{\Gamma(k_{j}\gamma_{j})}{\Gamma(k_{j}\gamma_{j}+n_{j,r})}  }      {U_{n_{j,1},\dots,n_{j,C_{0}}}(\max\bz_{j})}, ~~~~~k_{j}=\max\bz_{j},\dots,C_{0}, \label{eq: p(k|Z,W) 1}
\ee
where 
$U_{n_{j,1},\dots,n_{j,C_{0}}}(z) 
	= \sum_{k=z}^{C_{0}}  p_{0,j}(k)   \prod_{r=1}^{C_{0}}\{\Gamma(k\gamma_{j})/\Gamma(k\gamma_{j}+n_{j,r})\}$ and 
$n_{j,r}$ denotes the frequency of the $r\th$ category of the $j\th$ predictor $\bw_{j}$. 
By marginalizing out $\bpi_{\bk}$ and exploiting conditional independence relationships amongst different variables, 
we can show that the conditional distribution $p(\bk \mid \by,\bz,\bz^{\star},\blambda^{\star},\bpi^{\star})$ equals $p(\bk \mid \bz, \bw)=\prod_{j=1}^{q}p(k_{j} \mid \bz_{j},\bw_{j})$. 
This also follows easily by noting that the Markov blanket of $k_{j}$, after $\bpi_{\bk}$ have been integrated out, comprises precisely $\bz_{j}$ and $\bw_{j}$.
This allows us to design a collapsed Gibbs sampler that iterates between sampling $\bpi,\bz,\bz^{\star},\blambda^{\star}$ and $\bpi^{\star}$ from their closed form full conditionals, 
and then sampling $\bk$ from the closed form collapsed conditionals $p(\bk \mid \by,\bz,\bz^{\star},\blambda^{\star},\bpi^{\star})$.

To update the parameters $\blambda^{\star}$, $\bpi^{\star}$ and $\bz^{\star}$ of the DPM model, we used the approach of \cite{ishwaran_james:2001}, truncating the stick-breaking representation of the Dirichlet process prior on the mixture kernels $\blambda_{h_{1},\dots,h_{q}}$ at the $L\th$ component. 
In the examples that we considered in this article, the maximum number of categories was $4$. 
We set $L=100$, which sufficed for modeling conditional probability distributions supported on at most $4$ categories. 

We are now ready to describe our sampler. 
In what follows, $\bzeta$ denotes a generic variable that collects the variables that are not explicitly mentioned, including the data points $\by$. 
The sampler iterates between the following steps.


\begin{enumerate}
\item 
For each $(h_{1},\dots,h_{q})$, sample $z_{h_{1},\dots,h_{q}}^{\star}$ from its multinomial full conditional 
\bse
&&\hspace{-1cm} p(z_{h_{1},\dots,h_{q}}^{\star} = \ell \mid \bzeta)
\propto  \pi_{\ell}^{\star} \prod_{y=1}^{C_{0}} \{\lambda_{\ell}^{\star}(y)\}^{n_{h_{1},\dots,h_{q}}(y)}, 
\ese
where ${n}_{h_{1},\dots,h_{q}}(y) = \sum_{t=t^{\star}}^{T}1\{z_{1,t}=h_{1},\dots,z_{q,t}=h_{q},y_{t}=y\}$. 

\item 
For $\ell=1,\dots,L$, sample $V_{\ell}$ from their beta full conditionals 
\bse
 p(V_{\ell} \mid \bzeta) = \textstyle \Beta(1+n_{\ell}^{\star},\alpha_{0}+\sum_{k>\ell} n_{k}^{\star}),
\ese
where $n_{\ell}^{\star}=\sum_{(h_{1},\dots,h_{q})} 1\{z_{h_{1},\dots,h_{q}}^{\star}=\ell\}$,
and update $\bpi_{\ell}^{\star}$ accordingly.

\item 
For $\ell=1,\dots,L$, sample $\blambda_{\ell}^{\star}$ from their Dirichlet full conditionals 
\bse
&&\hspace{-1cm} \{\lambda_{\ell}^{\star}(1),\dots,\lambda_{\ell}^{\star}(C_{0})\} \mid \bzeta
\sim \Dir\left\{\alpha+{n}_{\ell}^{\star}(1),\dots,\alpha+{n}_{\ell}^{\star}(C_{0})\right\}, 
\ese
where ${n}_{\ell}^{\star}(y) = \sum_{(h_{1},\dots,h_{q})} 1\{z_{h_{1},\dots,h_{q}}^{\star}=\ell\}  {n}_{h_{1},\dots,h_{q}}(y)$.


\item
For $j=1,\dots,q$ and for $w_{j}=1,\dots,C_{0}$, sample
\bse
\{\pi_{1}^{(j)}(w_{j}),\dots,\pi_{k_{j}}^{(j)}(w_{j})\}\mid \bzeta \sim \Dir\{\gamma_{j}+{n}_{j,w_{j}}(1),\dots,\gamma_{j}+{n}_{j,w_{j}}(k_{j})\},
\ese
where $n_{j,w_{j}}(h_{j}) = \sum_{t=t^{\star}}^{T}1\{z_{j,t}=h_{j}, w_{j,t}=w_{j}\}$. 

\item
For $j=1,\dots,q$ and for $t=t^{\star},\dots,T$, sample the $z_{j,t}$'s from their multinomial full conditionals 
\bse
p(z_{j,t}=h\mid \bzeta,z_{\ell,t}=h_{\ell}, \ell \neq j) \propto \pi_{h}^{(j)}(w_{j,t}) \lambda_{z_{h_{1},\dots,h_{j-1},h,h_{j+1},\dots,h_{q}}^{\star}}^{\star}(y_{t}). 
\ese

\item Finally, for $j=1,\dots,q$, sample the $k_{j}$'s using their multinomial full conditionals 
\bse
&&\hspace{-1cm} p(k_{j} \mid \bzeta) = p(k_{j}\mid \bz_{j}, \bw_{j}) 
= \frac{  p_{0,j}(k_{j})\prod_{r=1}^{C_{0}}\frac{\Gamma(k_{j}\gamma_{j})}{\Gamma(k_{j}\gamma_{j}+n_{j,r})}  }      {U_{n_{j,1},\dots,n_{j,C_{0}}}(\max\bz_{j})}, ~~~~~k_{j}=\max\bz_{j},\dots,C_{0}. 
\ese
\end{enumerate}

The conditional probability $p(k_{j}\mid \bz_{j}, \bw_{j})$ depends on $\bz_{j}$ and $\bw_{j}$ only through $\max\bz_{j}$ and $n_{j,r}$, the frequencies of different categories of $w_{j}$.  
For a given data set, $n_{j,r}$'s are fixed quantities and $\max\bz_{j} \in \{1,\dots,C_{0}\}$. 
The values of $U_{n_{j,1},\dots,n_{j,C_{0}}}(z)$ for different possible values of $z$, and hence the distribution $p(k_{j}\mid \bz_{j}, \bw_{j})$, 
can thus be precomputed and stored before running the sampler. 

Using Stirling's approximation ${\Gamma(n+\alpha)}/{\Gamma(n)} \approx n^{\alpha}$, for moderately large  values of $n_{j,r}$, we can use 
\bse
&&\hspace{-1cm} p(k_{j}\mid \bz_{j}, \bw_{j})  \approx   \frac{  1  } {\sum_{\ell=\max\bz_{j}}^{C_{0}} \frac{p_{0,j}(\ell)}{p_{0,j}(k_{j})}  \prod_{r=1}^{C_{0}} n_{j,r}^{(k_{j}-\ell)\gamma_{j}} }, ~~~~~k_{j}=\max\bz_{j},\dots,C_{0}. 
\ese
To facilitate convergence, we initialize the component allocation variables $\bz$ at the cluster allocation values returned by an approximate two-stage sampler designed along the lines of \cite{yang_dunson:2015}, see Section \ref{sec: approximate sampler} in the Supplementary Materials.  
In experiments with synthetic and real data sets, $50,000$ MCMC iterations with the initial $10,000$ discarded as burn-in produced stable results, 
with trace plots and plots of running means and quantiles suggesting no convergence or mixing issues. 
To reduce autocorrelation, we thinned the post burn-in samples taking  every $5\th$ value.

\subsection{Testing} \label{sec: testing}
In many applications, it is of interest to test for the order of the Markov chain and the importance of a particular lag. The explosion in the number of parameters as the order increases and paucity of data in many applications have forced the literature on nonparametric tests of hypotheses to focus mostly on low order Markov chains 
\citep{avery_henderson:1999, quintana_newton:1998, xie_zimmerman:2014, besag_mondal:2013}.
A particularly attractive feature of our tensor factorization based approach is that many such hypotheses of interest can be expressed in terms of the variables $\bk$. 
For instance, 
the hypothesis that the $j\th$ lag is important is equivalent to $H_{0}: k_{j}>1$; 
the hypothesis that the chain is of maximal order $q_{0}$ translates to $H_{0}: k_{q_{0}}>1$ and $k_{j}=1$ for $j>q_{0}$; 
the hypothesis that the chain is of full order $q_{0}$ can be expressed as $H_{0}: k_{j}>1$ for $j=1,\dots,q_{0}$ and $k_{j}=1$ for all $j>q_{0}$
etc.

Following \cite{nobile:2004}, 
we use the number of non-empty components or clusters formed by the latent class allocation variables to estimate the number of mixture components. 
Denoting the number of clusters formed by $\bz_{j}$ by $\wt{k}_{j}$, 
we thus say that the $j\th$ lag is an important predictor of $y_{t}$ if and only if $\wt{k}_{j} > 1$.  
The hypotheses described above can be reformulated in terms of the $\wt{k}_{j}$'s accordingly. 
Clearly, $1 \leq \wt{k}_{j} \leq k_{j} \leq C_{0}$. 
The soft clustering aspect of our model makes it difficult to determine the induced prior on $\wt{k}_{j}$. 
However, the prior probabilities allocated to the different $H_{0}$'s described above depend only on the prior probabilities of the important special cases $\wt{k}_{j}=1$. 
Exploiting the symmetry of the Dirichlet prior on $\bpi_{k_{j}}^{(j)}$ and using equation (\ref{eq: p(ZIW)}) from \ref{appendix: posterior of k}, 
these probabilities can be easily obtained as 
\be
&& \hspace{-1cm} p_{0}(\wt{k}_{j}=1) = \sum_{k=1}^{C_{0}}p_{0j}(k) \sum_{\ell=1}^{k} p(z_{j,t}=\ell~\forall~t \mid k_{j}=k) 
= \sum_{k=1}^{C_{0}}p_{0j}(k)  k  \prod_{r=1}^{C_{0}} \left\{\frac{\Gamma(\gamma_{j}+n_{j,r})}{\Gamma(k\gamma_{j}+n_{j,r})}   \frac{\Gamma(k\gamma_{j})}{\Gamma(\gamma_{j})}\right\}     \nonumber \\
&& = \left\{\prod_{r=1}^{C_{0}}\gamma_{j}^{(n_{j,r})}\right\}    \left\{ \sum_{k=1}^{C_{0}} \frac{p_{0j}(k) k}{\prod_{r=1}^{C_{0}}(k\gamma_{j})^{(n_{j,r})}} \right\}.  \label{eq: p(ktilde=1)}
\ee
For moderately large values of $n_{j,r}$, Stirling's approximation can be used to obtain a simpler formula. 
For large values of $n_{j,r}$,  (\ref{eq: p(ktilde=1)}) will be close to $p_{0}(k_{j}=1)=p_{0j}(1)$. 

To conduct Bayesian tests for the different hypotheses described above, one may rely on the Bayes factor \citep{kass_raftery:1995} in favor of $H_{1}$ against $H_{0}$ given by
\bse
BF_{10} = \frac{p(H_{1} \mid \by) / p(H_{1})}  {p(H_{0} \mid \by) / p(H_{0})}, 
\ese
which can be easily estimated based on the output of the Gibbs sampler described in Section \ref{sec: posterior computation} 
with $p(H_{0} \mid \by)$ and $p(H_{1} \mid \by)$ equal to the proportion of samples in which the $\wt{k}_{j}$'s conform to $H_{0}$ and $H_{1}$, respectively. 
Results of simulation experiments evaluating performance of the Bayes factor based tests are summarized in Section \ref{sec: simulation experiments}. 

\section{Simulation Experiments} \label{sec: simulation experiments}
We designed simulation experiments to evaluate the performance of our method in estimating various aspects of the transition dynamics in a wide range of scenarios. 
Some of the cases were generated to closely mimic the real data sets that we analyzed in Section \ref{sec: applications}. 
We consider the cases 
(A) $[4, \{1,2,3\}]$, 
(B) $[3, \{1,2,3\}]$, 
(C) $[4, \{1,2,4\}]$,  
(D) $[3, \{1,2,4\}]$,  
(E) $[4, \{1,3,5\}]$,    
(F) $[3, \{1,3,5\}]$,    
(G) $[3, \{1,4,8\}]$,  and
(H) $[2, \{1,4,8\}]$,
where $[C_{0}, \{i_{1},\dots,i_{r}\}]$ means that the sequence has $C_{0}$ categories and $\{y_{t-i_{1}},\dots,y_{t-i_{r}}\}$ are the true important lags.  
In each case, we considered two sample sizes $T=200, 500$ and generated an additional $N=500$ test data points to evaluate prediction performance. 
The maximal order $q$ of the chain was chosen to be two more than the most distant important lag, namely $q=5,6,7$ and $10$, respectively. 
To generate the true transition probability tensors, for each combination of the true lags, 
we first generated the probability of the first response category as $f(U_{1})=U_{1}^{2}/\{U_{1}^{2}+(1-U_{1})^{2}\}$ with $U_{1} \sim \Unif(0,1)$. 
The probabilities of the remaining categories are then generated via a stick-breaking type construction as $f(U_{2}) \{1-f(U_{1})\}$ with $U_{2}\sim \Unif(0,1)$ and so on, 
until the next to last category $(C_{0}-1)$ is reached. 
The hyper-parameters were set at $\alpha=1/C_{0}$, $\alpha_{0}=1$ and $\gamma_{j}=1/C_{0}$ for all $j$. 
We prescribe using $\varphi=1/2$, which produced good results in synthetic data sets and real applications, as a default value for $\varphi$.
The reported results are based on $100$ simulated data sets in each case. 
We coded in MATLAB. 
For the case (G) described above, with $C_{0}=3$ categories and $T=500$ data points, $50,000$ MCMC iterations required approximately $30$ minutes on an ordinary laptop.

We compared our approach with a multinomial logit model that includes the lags of order up to $q$ as linear predictors and ignores interactions, 
a variable length Markov chain (VLMC) model, 
a sparse Markov chain (SMC) model, 
and a mixture transition distribution (MTD) model. 
We also included a simple random forest \citep{breiman:2001} based model (RFMC) which, like VLMC, is also tree-based but, unlike VLMC, does not enforce a strict top-down search.
The multinomial logit model and the RFMC model were implemented using respectively the VGAM \citep{yee:2010} and the randomForest \citep{liaw_weiner:2002} packages in R. 
The VLMC model was implemented using the R package VLMC with the pruning parameter selected using the AIC criterion \citep{machler_buhlmann:2004}.  
The SMC and the MTD models were implemented using MATLAB codes downloaded 
from http://www.helsinki.fi/bsg/filer/SMCD.zip and http://lib.stat.cmu.edu/matlab/GMTD, respectively. 
Instead of refitting the MTD and the SMC models with different possible choices for the maximal order, 
we set the maximal order at the corresponding true value, giving these models an undue advantage over others.  

Performance in estimating the transition probabilities and predicting one step ahead response values are summarized in Table \ref{tab: aL1} and Table \ref{tab: CLER 1}, respectively. 
The average $L_{1}$ errors were estimated by $\sum_{t=T+1}^{T+N} \sum_{y=1}^{C_{0}} |P_{0}(y\mid y_{t-1},\dots,y_{t-q})-\wh{P}(y\mid y_{t-1},\dots,y_{t-q})|/(C_{0}N)$, where $P_{0}$ and $\wh{P}$ are the true and the estimated transition probability tensors, respectively. 
The proposed model performed competitively with VLMC and SMC approaches when the maximal orders were small and there were no gaps in the set of important lags. 
When the true maximal orders were increased and lag gaps were introduced, the proposed approach vastly outperformed all competitors. 
In the latter cases, the VLMC method, which employs a tree based top-down approach to determine serial dependencies, 
fails to eliminate the unimportant intermediate lags leading to its poor performance. 
The SMC and the RFMC methods can accommodate lag gaps and in these cases their performances were generally superior to that of the VLMC model. 
However, their strategy of hard clustering large conditioning sequences becomes increasingly ineffective as the true maximal order increases 
and the proposed conditional tensor factorization based approach starts to vastly dominate. 
It is to be noted that our implementation of the SMC method assumed the true maximal order to be known in each case. 
The approach outlined in \cite{xiong_etal:2015} to determine the optimal order did not work well in our experiments, producing significantly worse results. 
The MTD approach had poor performance in all cases, likely due to its restrictive assumption of simple additive effects of different lags.

\begin{table}[ht!]
\begin{center}
\begin{tabular}{|c|c|c c c c c c|}
\hline
\multirow{2}{*}{Truth} 			& \multirow{2}{*}{Sample Size}	& \multicolumn{6}{|c|}{Average $L_{1}$ Error $\times 100$} \\ \cline{3-8}
							&		& MLGT		& VLMC		& SMC 		& RFMC	& MTD		& CTF	  \\  \hline\hline
\multirow{2}{*}{(A) $4, \{1,2,3\}$}	& 200	& 20.14		& 11.59		& \bf{9.77}		& 12.54	& 20.45		& {12.67}       \\
							& 500	& 19.99		& 7.31		& \bf{6.95}		& 10.70	& 20.16		& {7.66}        \\\hline
\multirow{2}{*}{(B) $3, \{1,2,3\}$}	& 200	& 20.60		& 9.24		& \bf{8.54}		& 12.59	& 22.14		& {11.12}       \\
							& 500	& 19.74		& \bf{5.07}		& 5.21		& 10.59	& 21.07		& {5.56}        \\\hline\hline

\multirow{2}{*}{(C) $4, \{1,2,4\}$}	& 200	& 20.19		& 16.24		& 14.66	& 14.42	& 20.84		& \bf{13.81}       \\
							& 500	& 19.40		& 11.09		& 9.61	& 11.35	& 20.10		& \bf{7.79}        \\\hline
\multirow{2}{*}{(D) $3, \{1,2,4\}$}	& 200	& 22.05		& 13.59		& 11.22	& 13.41	& 23.45		& \bf{11.09}       \\
							& 500	& 21.38		& 8.33		& 7.26	& 11.03	& 22.74		& \bf{6.12}        \\\hline\hline

\multirow{2}{*}{(E) $4, \{1,3,5\}$}	& 200	& 21.39		& 20.51		& 18.93	& 16.62 	& 21.56		& \bf{15.30}        \\
							& 500	& 20.74		& 16.66		& 14.89	& 13.31	& 20.91		& \bf{8.24}   	\\\hline
\multirow{2}{*}{(F) $3, \{1,3,5\}$}	& 200	& 23.57		& 18.83		& 16.12	& 15.67	& 24.46		& \bf{11.58}        \\
							& 500	& 22.69		& 13.18		& 11.24	& 12.05	& 23.64		& \bf{6.38}   	\\\hline\hline

\multirow{2}{*}{(G) $3, \{1,4,8\}$}	& 200	& 23.86		& 24.92		& 26.82 	& 18.24	& 24.42		& \bf{12.00}   	\\
							& 500	& 23.41		& 22.33		& 24.04	& 14.41	& 24.18		& \bf{6.60}   	\\\hline
\multirow{2}{*}{(H) $2, \{1,4,8\}$}	& 200	& 19.16		& 20.14		& 17.53	& 13.31	& 21.80		& \bf{7.17}   	\\
							& 500	& 16.69		& 13.07		& 11.07	& 9.99	& 19.74		& \bf{3.62}   	\\\hline
\end{tabular}
\caption{\baselineskip=10pt Average $L_{1}$ distances between the true and the estimated transition probability tensors for our conditional tensor factorization (CTF) based approach 
compared with 
a multinomial logit model (MLGT), 
a variable length Markov chain (VLMC) model, 
a sparse Markov chain (SMC) model, 
a random forest based (RFMC) model, 
and a mixture transition distribution (MTD) model. 
In the first column, $C_{0}, \{i_{1},\dots,i_{r}\}$ means that the sequence has $C_{0}$ categories and $\{y_{t-i_{1}},\dots,y_{t-i_{r}}\}$ are the true important lags.  
See Section \ref{sec: simulation experiments} for additional details.
The minimum value in each row is highlighted.
}
\label{tab: aL1}
\end{center}
\end{table}

\begin{table}[ht!]
\begin{center}
\begin{tabular}{|c|c|c c c c c c|}
\hline
\multirow{2}{*}{Truth} 			& \multirow{2}{*}{Sample Size}	& \multicolumn{6}{|c|}{Classification Error Rates $\times 100$} \\ \cline{3-8}
							&		& MLGT		& VLMC		& SMC 		& RFMC	& MTD		& CTF	  \\  \hline\hline
\multirow{2}{*}{(A) $4, \{1,2,3\}$}	& 200	& 50.39		& 36.50		& \bf{33.45}	& 35.82	& 51.40		& {35.37}       \\
							& 500	& 50.73		& 31.27		& \bf{29.83}	& 33.39	& 51.54		& {30.05}        \\\hline
\multirow{2}{*}{(B) $3, \{1,2,3\}$}	& 200	& 40.28		& 26.42		& \bf{24.53}	& 27.51	& 42.53		& {26.75}       \\
							& 500	& 37.86		& 22.43		& 22.88 		& 23.30	& 39.84		& \bf{21.99}        \\\hline\hline

\multirow{2}{*}{(C) $4, \{1,2,4\}$}	& 200	& 50.52		& 44.03		& 42.36		& 39.47	& 51.23		& \bf{37.05}       \\
							& 500	& 48.22		& 35.14		& 35.53		& 32.87	& 49.26		& \bf{28.95}        \\\hline
\multirow{2}{*}{(D) $3, \{1,2,4\}$}	& 200	& 41.67		& 30.28		& 27.14		& 28.69	& 43.94		& \bf{25.19}       \\
							& 500	& 40.70		& 25.22		& 23.82 		& 24.70	& 43.03		& \bf{22.07}        \\\hline\hline

\multirow{2}{*}{(E) $4, \{1,3,5\}$}	& 200	& 52.76		& 51.28		& 42.36		& 42.33	& 52.88		& \bf{38.35}        \\
							& 500	& 50.77		& 44.95		& 38.38		& 37.43	& 50.98		& \bf{29.90}   	\\\hline
\multirow{2}{*}{(F) $3, \{1,3,5\}$}	& 200	& 45.26		& 37.62		& 32.39		& 31.70	& 45.59		& \bf{25.59}        \\
							& 500	& 42.88		& 30.26		& 27.28		& 26.11	& 43.78		& \bf{22.25}   	\\\hline\hline

\multirow{2}{*}{(G) $3, \{1,4,8\}$}	& 200	& 45.95		& 47.23		& 47.27 		& 35.36	& 45.99		& \bf{26.86}   	\\
							& 500	& 45.00		& 43.68		& 43.59		& 29.78	& 46.11		& \bf{22.90}   	\\\hline
\multirow{2}{*}{(H) $2, \{1,4,8\}$}	& 200	& 25.23		& 26.78		& 23.97		& 18.77	& 26.93		& \bf{14.55}   	\\
							& 500	& 22.37		& 19.89		& 18.42		& 15.02	& 24.88	& \bf{13.78}   	\\\hline
\end{tabular}
\caption{\baselineskip=10pt Classification error rates of the conditional tensor factorization (CTF) based approach 
in predicting one step ahead response values compared with 
a variable length Markov chain (VLMC) model, 
a sparse Markov chain (SMC) model, 
a random forest based (RFMC) model, 
and a mixture transition distribution (MTD) model. 
IIn the first column, $C_{0}, \{i_{1},\dots,i_{r}\}$ means that the sequence has $C_{0}$ categories and $\{y_{t-i_{1}},\dots,y_{t-i_{r}}\}$ are the true important lags.  
See Section \ref{sec: simulation experiments} for additional details.
The minimum value in each row is highlighted.
}
\label{tab: CLER 1}
\end{center}
\end{table}

\begin{figure}[ht!]
\begin{center}
        \includegraphics[height=10cm,width=15cm, trim=3cm 9cm 3cm 9cm]{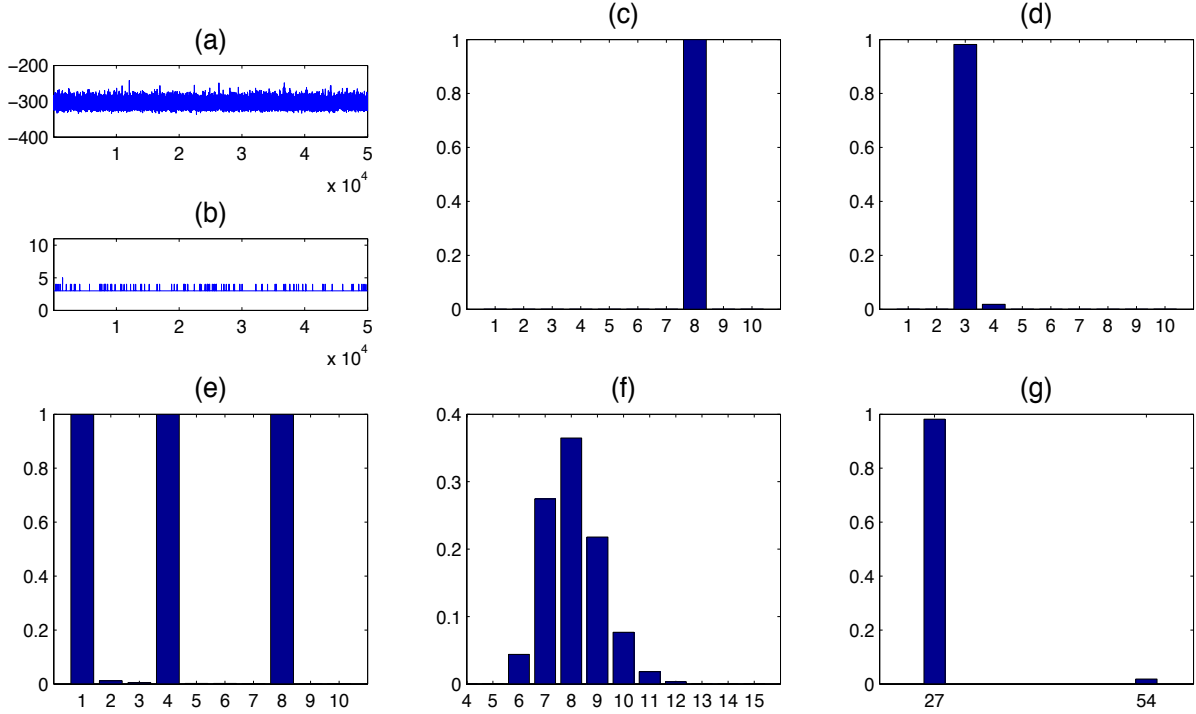}
        \caption{Results of simulation experiments for the case (G) with $C_{0}=3$ categories, true important lags $\{y_{t-1}, y_{t-4}, y_{t-8}\}$ and sample size $T = 500$ for the data set with the median classification error rate. 
         (a) trace plot of marginal likelihood $p(\by\mid \bz)$; 
         (b) trace plot of the number of important lags; 
         (c) relative frequency distribution of the maximal order; 
         (d) relative frequency distribution of the number of important lags; 
         (e) inclusion proportions of different lags; 
         (f) relative frequency distribution of the number of clusters of the probability kernels $\blambda_{h_{1},\dots,h_{q}}$; and   
         (g) relative frequency distribution of $\prod_{j=1}^{q}k_{j}$, the number of possible combinations of $(h_{1},\dots,h_{q})$. 
        Panels (c)-(g) are based on thinned samples after burn-in. 
        See Section \ref{sec: HOMC via CTF} and Section \ref{sec: simulation experiments} for additional details.  
        }
        \label{fig: Simulation Experiments 1: MCMC Output}
\end{center}
\end{figure}

\begin{figure}[ht!]
\begin{center}
        \includegraphics[height=8cm,width=15cm,trim=3cm 9cm 3cm 10cm]{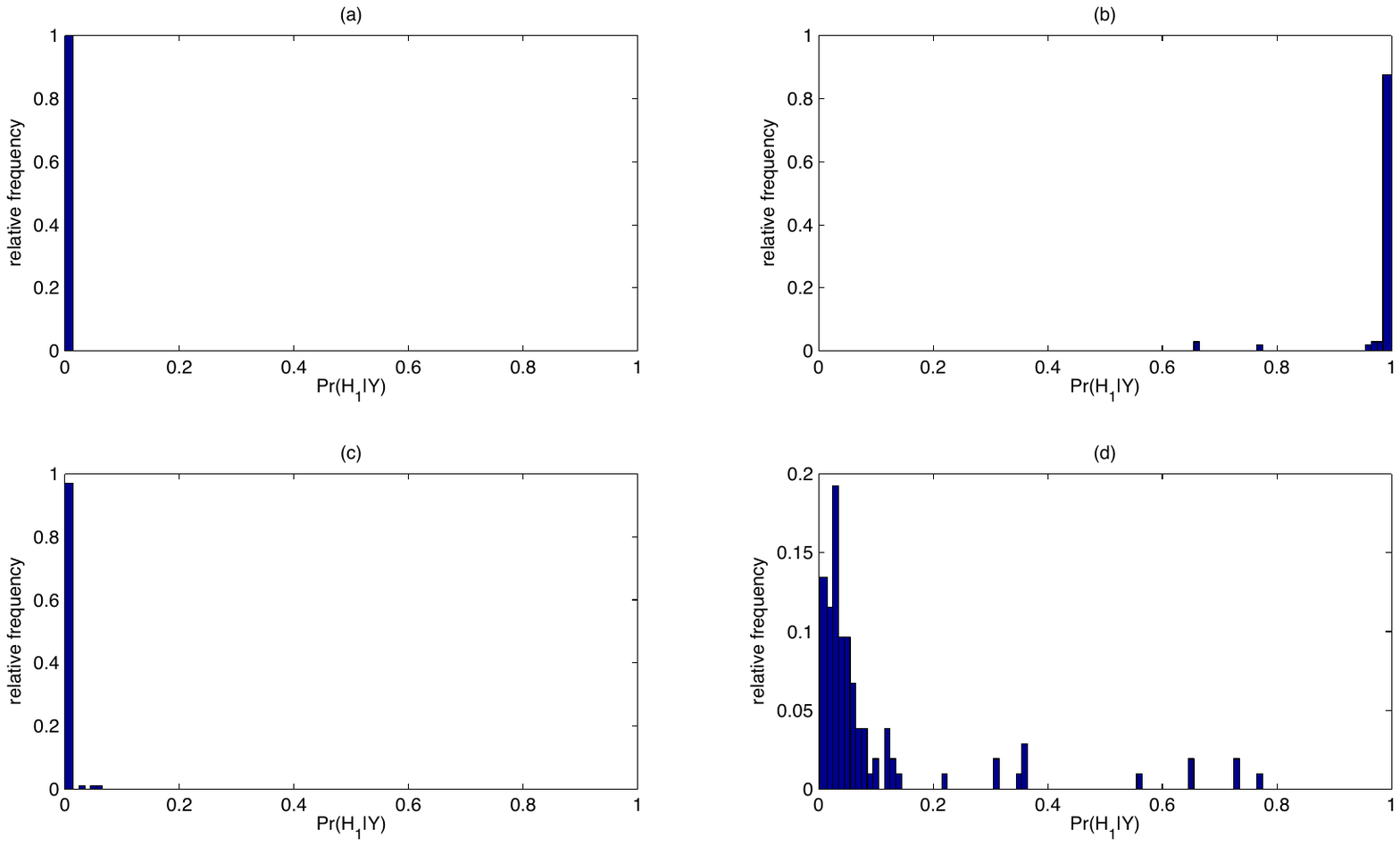}
        \caption{Histograms of posterior probabilities of alternative hypotheses $H_{1}$ based on $100$ simulated data sets for the case (G) with $C_{0}=3$ categories, true important lags $\{y_{t-1}, y_{t-4}, y_{t-8}\}$ and sample size $T = 500$. 
        (a)  $H_{0}:$ the lag $y_{t-4}$ is important; 
        (b) $H_{0}:$ the lag $y_{t-5}$ is important; 
        (c) $H_{0}:$ $y_{t-8}$ is an important lag but $y_{t-9}$ and $y_{t-10}$ are not, that is, the chain is of maximal order $8$; and 
        (d) $H_{0}:$ the only important lags are $\{y_{t-1},y_{t-4}, y_{t-8}\}$. 
        For the cases (a), (c) and (d), the corresponding $H_{0}$'s were actually true, whereas for the case (b), $H_{0}$ was false. 
        See Sections \ref{sec: testing} and \ref{sec: simulation experiments} for additional details. 
        }
        \label{fig: Simulation Experiments 2: Testing}
\end{center}
\end{figure}

Figure \ref{fig: Simulation Experiments 1: MCMC Output} summarizes the results produced by our method for the case (G) $[3, \{1,4,8\}]$ with $T=500$ data points 
for the data set corresponding to the median classification error rate. 
Panels (c)-(e) in Figure \ref{fig: Simulation Experiments 1: MCMC Output} illustrate the method's ability to identify the important lags and the maximal order of the chain. 

To assess testing performance, we considered the hypotheses 
(a) $H_{0}: \wt{k}_{4}>1$, 
(b) $H_{0}: \wt{k}_{5}>1$, 
(c): $H_{0}: \wt{k}_{8}>1$ and $\wt{k}_{9}=\wt{k}_{10}=1$, and 
(d) $H_{0}: \wt{k}_{j}>1$ for $j=1,4,8$ and $\wt{k}_{j}=1$ otherwise
for the case (G) $[3, \{1,4,8\}]$ described above with $500$ data points. 
Figure \ref{fig: Simulation Experiments 2: Testing} shows histograms of the estimated posterior probabilities of the alternative hypotheses based on $100$ simulated data sets. 
For the cases (a), (c) and (d), when the corresponding $H_{0}$'s are actually true, the method appropriately assigns values close to zero, whereas for the case (b), when $H_{0}$ is actually false, the estimated posterior probabilities are very close to one. 


\section{Applications} \label{sec: applications}
In this section, we discuss two applications of the proposed conditional tensor factorization approach. 
In each case, we set the maximal possible order at $q=10$. 
In experiments with higher values of $q$, the results remained practically unchanged. 
An additional application of the procedure described in Section \ref{sec: testing} to test the order of serial dependence in a DNA sequence is presented in Section \ref{sec: human gene data set} of the Supplementary Materials, showing substantially improved results relative to competitors.
These data sets have all been analyzed previously in the literature but our proposed nonparametric approach provides new insights into their serial dependence structures. 
Results produced by the VLMC and the RFMC methods for these data sets are deferred to Section \ref{sec: vlmc & rfmc summary} of the Supplementary Materials.  

\subsection{Epileptic Seizure Data Set}
We first reanalyze a data set from \cite{berchtold_raftery:2002} (BR), originally presented in \cite{mcdonald_zucchini:1997}. 
The data set comprises a binary time series describing whether a patient experienced epileptic seizures on $204$ consecutive days. 
The two states correspond to either no epileptic seizure or at least one epileptic seizure. 
BR analyzed the data set using the MTD model and found that $y_{t}$ is best explained by an MTD model with $8$ lags with the lag $y_{t-8}$ being the most important one. 
The other important lags were $y_{t-1}, y_{t-4}, y_{t-5}$ and $y_{t-7}$. 

\begin{figure}[ht]
\begin{center}
        \includegraphics[height=10cm,width=14cm, trim=3cm 9cm 3cm 9cm]{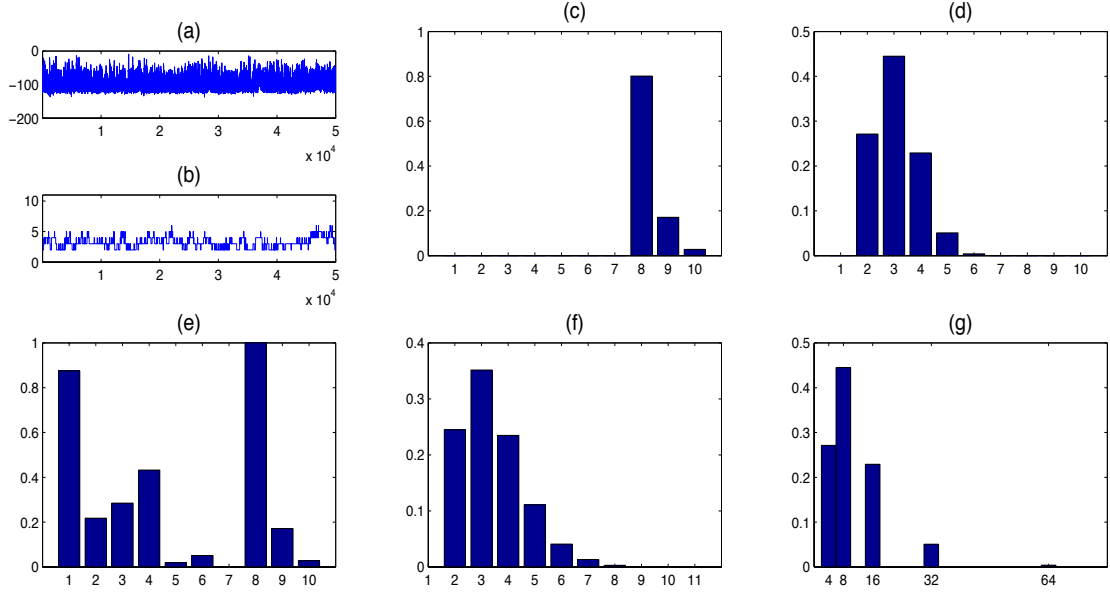}
        \caption{Results for the seizure data set. 
         (a) trace plot of marginal likelihood $p(\by\mid \bz)$; 
         (b) trace plot of the number of important lags; 
         (c) relative frequency distribution of the maximal order; 
         (d) relative frequency distribution of the number of important lags; 
         (e) inclusion proportions of different lags; 
         (f) relative frequency distribution of the number of clusters of the probability kernels $\blambda_{h_{1},\dots,h_{q}}$; and   
         (g) relative frequency distribution of $\prod_{j=1}^{q}k_{j}$, the number of possible combinations of $(h_{1},\dots,h_{q})$. 
        Panels (c)-(g) are based on thinned samples after burn-in. 
        See Section \ref{sec: applications} for additional details.  
        }
        \label{fig: Application 1: Seizure Data Set}
\end{center}
\end{figure}

Figure \ref{fig: Application 1: Seizure Data Set} summarizes the results obtained by our conditional tensor factorization approach. 
In agreement with BR, our analysis provides strong evidence for a Markov chain of maximal order $8$ with $y_{t-8}$ being the most important lag with an inclusion probability of one. 
With an inclusion probability close to $0.85$, $y_{t-1}$ was the second most important predictor. 
However, in contrast with BR, the distribution of the number of important lags was concentrated around $3$ with the lags $\{y_{t-1},y_{t-4},y_{t-8}\}$ appearing together the maximum number of times. 
In particular, the inclusion probabilities of $y_{t-5}$ and $y_{t-7}$ were very close to zero suggesting that these lags were not important predictors of $y_{t}$.

\subsection{Song of the Wood Pewee Data Set} \label{sec: wood pewee}
Next, we reanalyze a data set from \cite{raftery_tavare:1994} (RT) 
that describes the morning song of the wood pewee, a North American song bird, comprising $3$ distinct phrases, labeled $1,2$ and $3$. 
An interesting feature of the data set is that the series is dominated by two repeating patterns, namely $1312$ and $112$. 
The repeating patterns indicate strong interactions among the lags and MTD models are not suitable for such data sets. 
As pointed out by RT, although the first pattern is of length $4$, it can be specified by four transitions of order $2$, namely $1|31, 3|12, 1|21, 2|13$. 
Likewise, the second repeating pattern $112$ can be defined by the second order transitions $1|12, 1|21, 2|11$. 
The transition $1|21$ and the conditioning sequence $12$ appear in both patterns. 
To accommodate these features, RT modeled the transition probabilities as
\bse
p(y_{t}=y_{t}\mid y_{t-1}=y_{t-1},y_{t-2}=y_{t-2}) = \left\{\begin{array}{ll}
\alpha_{h} 		& \text{if}~\{y_{t-1},y_{t-2}\} \in A_{h}, \\
(1-\alpha_{h})\frac{\pi_{y_{t}}}{\sum_{\{y:(y\mid y_{t-1},y_{t-2})\notin A_{h}\}}\pi_{j}} 		& \text{if}~\{y_{t-1},y_{t-2}\} \in B_{h}, \\
\gamma_{y_{t}}		& \text{if}~\{y_{t-1},y_{t-2}\}=\{1,2\}, \\
\pi_{y_{t}}			& \text{otherwise},
\end{array}\right.
\ese
where $h=1,2$, $A_{1}=\{1|31,1|21,2|13\}$, $A_{2}=\{2|11\}$, $B_{h}=\{(\wt{y}_{t}\mid \wt{y}_{t-1}, \wt{y}_{t-2}):  (\wt{y}_{t}\mid \wt{y}_{t-1}, \wt{y}_{t-2})\notin A_{h} ~ \text{but there exists}~(y_{t}\mid y_{t-1},y_{t-2})~\text{with}~y_{t-1}=\wt{y}_{t-1}~\text{and}~y_{t-2}=\wt{y}_{t-2}\}$, $1\leq \alpha_{h}\leq 1, 0 \leq \pi_{y} \leq 1, \sum_{y}\pi_{y}=1$ and $0 \leq \gamma_{y} \leq 1, \sum_{y}\gamma_{y}=1$. 
The construction of such complex models with different parameterizations for different conditioning sequences requires critical understanding of the important features of the transition dynamics on a case by case basis and can not be easily generalized.

\begin{figure}[ht]
\begin{center}
        \includegraphics[height=10cm,width=14cm, trim=3cm 9cm 3cm 9cm]{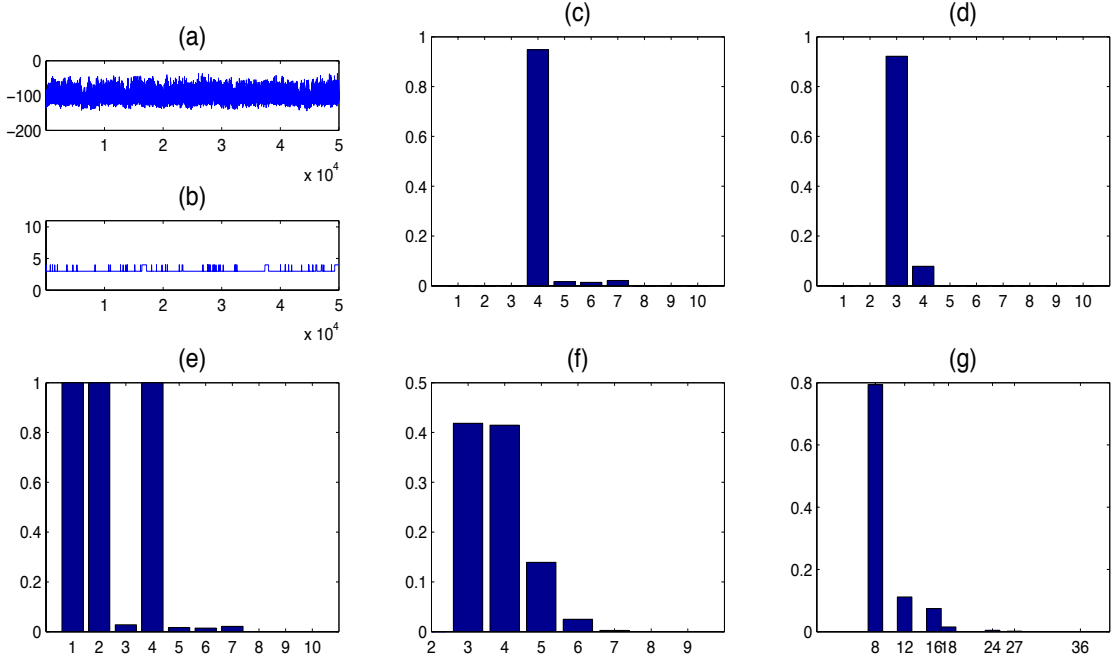}
        \caption{Results for the wood pewee data set. 
         (a) trace plot of marginal likelihood $p(\by\mid \bz)$; 
         (b) trace plot of the number of important lags; 
         (c) relative frequency distribution of the maximal order; 
         (d) relative frequency distribution of the number of important lags; 
         (e) inclusion proportions of different lags; 
         (f) relative frequency distribution of the number of clusters of the probability kernels $\blambda_{h_{1},\dots,h_{q}}$; and   
         (g) relative frequency distribution of $\prod_{j=1}^{q}k_{j}$, the number of possible combinations of $(h_{1},\dots,h_{q})$. 
        Panels (c)-(g) are based on thinned samples after burn-in. 
        See Section \ref{sec: applications} for additional details.  
        }
        \label{fig: Application 2: Wood Pewee Data Set}
\end{center}
\end{figure}

Figure \ref{fig: Application 2: Wood Pewee Data Set} summarizes the results obtained by applying our conditional tensor factorization approach to the first $500$ data points of the wood pewee data set. 
Panels (c), (d) and (e) of Figure \ref{fig: Application 2: Wood Pewee Data Set} indicate strong evidence of a Markov chain of maximal order $4$ with three important lags $\{y_{t-1}, y_{t-2}, y_{t-4}\}$. 
The inclusion proportions of these three lags were all close to one, whereas the inclusion proportion of the intermediate lag $y_{t-3}$ was close to zero. 
This suggests that given $\{y_{t-1}, y_{t-2}, y_{t-4}\}$, $y_{t-3}$ carries little additional information useful for predicting $y_{t}$.  

The results can be explained by first noting that a third order representation of the two repeating patterns $1312$ and $112$ comprise the transitions $1|312, 3|121,1|213,2|131$ and $1|121,1|211,2|112$, respectively, 
with the conditioning sequence $121$ appearing in both sets of transitions. 
A fourth order representation, on the contrary, gives transitions with unique conditioning sequences, namely $1|3121,3|1213,1|2131,2|1312$ and $1|1211,1|2112,2|1121$, respectively. 
Also, if the third lag $y_{t-3}$ is dropped, we still obtain transitions with unique conditioning sequences, 
namely $1|31\cdot1,3|12\cdot3,1|21\cdot1,2|13\cdot2$ and $1|12\cdot1,1|21\cdot2,2|11\cdot1$, respectively. 
It is thus clear that a Markov chain of maximal order $4$ with three important lags $\{y_{t-1},y_{t-2},y_{t-4}\}$ would provide a good characterization of the transition dynamics of the wood pewee data set, 
as is captured by the proposed tensor factorization based approach.

\section{Discussion} \label{sec: discussion} 
The proposed nonparametric Bayesian method provides a flexible yet parsimonious representation of higher order Markov chains, allowing automated identification of the set of important lags 
while also facilitating testing of many hypotheses of practical interest.   
In simulation experiments, our method substantially out-performed competitors when there were gaps in the set of important lags, while performing competitively with existing methods in other cases. 
We have found that such gaps are commonplace in applications we have considered.

While the focus of this paper has been on higher order homogeneous Markov models, 
the proposed methodology can be easily extended to nonhomogeneous cases in which the transition dynamics is also influenced by exogenous predictors. 
Indeed, our computer codes already accommodate sequentially varying categorical predictors. 
Multiple categorical sequences can also be easily accommodated with the sequence label treated as an exogenous sequence specific categorical predictor. 
Additional important directions of ongoing research include extensions of the methodology to other discrete state space dynamical systems, including higher order hidden Markov models and models for spatial and spatio-temporal categorical data sets.

\baselineskip=17pt
\section*{Supplementary Materials}
The Supplementary Materials present 
some additional figures,
describe the approximate two-stage sampler used to determine the starting values of the MCMC sampler, 
discuss MCMC diagnostics, 
present an additional application of the proposed methodology,
and summarize the results produced by the VLMC and the RFMC methods for the data sets discussed in Section \ref{sec: applications}. 


\newcommand{\Appendix}{\appendix\def\thesection{Appendix~\Alph{section}}\def\thesubsection{\Alph{section}.\arabic{subsection}}}
\section*{Appendix}
\begin{appendix}
\Appendix
\renewcommand{\theequation}{A.\arabic{equation}}
\setcounter{equation}{0}
\baselineskip=18pt

\section{Proof of Theorem 1} \label{appendix: consistency}
Consider a Markov chain $\{y_{t}\}$ of maximal order $q$ with finite state space $\Y$ and transition probability tensor $P$. 
Then $\{\bs_{t}=(y_{t},\dots,y_{t-q+1})\trans\}$ is a first-order Markov chain with state space $\S=\Y^{q}$
and transition probability matrix $\wt{P}$ with entries  
\bse
&&\hspace{-1cm} \wt{P}[(j_{t-1},\dots,j_{t-q}), (i_{t},\dots,i_{t-q+1})  ] \\
&& = \Pr[(y_{t}=i_{t},\dots,y_{t-q+1}=i_{t-q+1}) \mid (y_{t-1}=j_{t-1},\dots,y_{t-q}=j_{t-q}) ] \\
&& = 
\left\{\begin{array}{ll}
P(y_{t}=i_{t}\mid y_{t-1}=j_{t-1},\dots,y_{t-q}=j_{t-q}),  & \text{if}~i_{t-\ell}=j_{t-\ell}~\text{for}~\ell=1,\dots,(q-1),\\
0,	& \text{otherwise}.
\end{array}\right.
\ese
The following lemma establishes a general posterior consistency result for first order Markov chain models. 
With $\wt{P}$ and $\wt{P}_{0}$ denoting the transition probability matrices of the first order representations of two Markov chains with respective transition probability tensors $P$ and $P_{0}$, 
we have $d(P,P_{0})=\sum_{\bs_{1}}\sum_{\bs_{2}}|\wt{P}(\bs_{1},\bs_{2})-\wt{P}_{0}(\bs_{1},\bs_{2})|$.
The conclusion of Theorem \ref{Thm: Main Theorem 0} thus follows as a consequence. 
\begin{Lem}
Let $\{y_{t}\}$ be an ergodic Markov chain with finite state space and transition probability matrix $P \in \calP$. 
Let $\Pi$ be a prior on $\calP$. 
Then for any $P_{0}$ in the Kullback-Leibler support of $\Pi$ 
and any $\delta>0$,
$\Pi\left\{P: \textstyle d(P,P_{0})>\delta \mid \by_{1:T}\right\} \to 0 ~\text{a.s.}~P_{0}$. 
\end{Lem}
\begin{Proof*}
Let $\Y$ denote the state space of $\{y_{t}\}$. 
Since $\{y_{t}\}$ is ergodic, it has a unique stationary distribution $\pi_{0}$, with $\pi_{0}(j)>0$ for any $j\in \Y$. 
We define the empirical stationary distribution as $\wh{\pi}_{T}(j)=\sum_{t=1}^{T}1\{y_{t}=j\}/T=n_{j}/T$ for any $j\in\Y$. 
Likewise, for any $i,j\in \Y$, we define the empirical transition probability matrix as $\wh{P}_{T}(i,j)=\sum_{t=1}^{T}1\{y_{t-1}=i,y_{t}=j\}/\sum_{t=1}^{T}1\{y_{t}=i\} = n_{i,j}/n_{j}$. 
Define $d_{\pi_{0}}(P,P_{0})=\sum_{i}\sum_{j}\pi_{0}(i)|P(i,j)-P_{0}(i,j)|$ and $K_{\pi_{0}}(P,P_{0})=\sum_{i}\sum_{j}\pi_{0}(i)P_{0}(i,j) \log\frac{P(i,j)}{P_{0}(i,j)}=K(\pi_{0}P,\pi_{0}P_{0})$. 
Also let $V=\{P: d_{\pi_{0}}(P,P_{0})>\delta\min_{i}\pi_{0}(i)\}$. 
Then $\{P: d(P,P_{0})>\delta\}\subseteq V$. 
We have 
\be
\Pi\left\{P: d(P,P_{0})>\delta \mid \by_{1:T}\right\} \leq \Pi(V \mid \by_{1:T}) = \frac{\int_{V} \exp\left\{-T   \sum_{i}\sum_{j}   \frac{n_{i}}{T}   \frac{n_{i,j}}{n_{i}}     \log \frac{P_{0}(i,j)}{P(i,j)}\right\}  d\Pi(P)}{\int_{\calP} \exp\left\{-T   \sum_{i}\sum_{j}  \frac{n_{i}}{T}    \frac{n_{i,j}}{n_{i}}     \log \frac{P_{0}(i,j)}{P(i,j)}\right\}  d\Pi(P)}. \nonumber\\       \label{eq1: proof}
\ee
By ergodic theorem, $\wh{\pi}_{T}$ and $\wh{P}_{T}$ converge almost surely to $\pi_{0}$ and $P_{0}$, respectively \citep{eichelsbacher_ganesh:2002}. 
Therefore, for the numerator, we have 
\bse
&&\hspace{-1cm} \lim_{T\to\infty}\sum_{i}\sum_{j}  (n_{i}/T)  (n_{i,j}/n_{i})  \log \{P_{0}(i,j)/P(i,j)\} \\
&&= \lim_{T\to\infty} \sum_{i}\sum_{j}  (n_{i}/T)  (n_{i,j}/n_{i})  \log \{(n_{i,j}/n_{i})/P(i,j)\} 
= \lim_{T\to\infty} K(\wh\pi_{T}\wh{P}_{T},\wh\pi_{T} P). 
\ese
For any $P\in V$ and $T$ sufficiently large, we have 
\bse
&&\hspace{-1cm} K(\wh\pi_{T}\wh{P}_{T},\wh\pi_{T} P) \geq d_{\wh{\pi}_{T}}^{2}(\wh{P}_{T},P)/4 \\
&& \geq \{d_{\wh{\pi}_{T}}(P,P_{0})-d_{\wh{\pi}_{T}}(\wh{P}_{T},P_{0})\}^{2}/4  \geq (2\delta/3-\delta/3)^{2}/4=\delta^{2}/36 ~~ \text{a.s.}~P_{0}. 
\ese
Therefore, with $\beta<\delta^{2}/36$, we have 
\be
\lim_{T\to\infty}\exp(\beta T) \int_{V} \exp\left\{-T   \sum_{i}\sum_{j}  \frac{n_{i}}{T}    \frac{n_{i,j}}{n_{i}}     \log \frac{P_{0}(i,j)}{P(i,j)}\right\}  d\Pi(P) \nonumber\\
\leq \lim_{T\to\infty}\exp\{(\beta-\delta^{2}/36)T\}  
= 0 ~~ \text{a.s.}~P_{0}.  \label{eq2: proof}
\ee
Similarly, for the denominator, we have $- \sum_{i}\sum_{j}   (n_{i}/T)   (n_{i,j}/n_{i})     \log \{P_{0}(i,j)/P(i,j)\} \to -K_{\pi_{0}}(P_{0},P)>-\epsilon$ a.s. $P_{0}$ for any $P$ with $K_{\pi_{0}}(P_{0},P)<\epsilon$. 
For any $\beta>0$, choosing $\epsilon=\beta/2$, using Fatou's lemma we have 
\be
\exp(\beta T) \int_{\calP} \exp\left\{-T   \sum_{i}\sum_{j}  \frac{n_{i}}{T}    \frac{n_{i,j}}{n_{i}}     \log \frac{P_{0}(i,j)}{P(i,j)}\right\}  d\Pi(P) 
\to \infty ~~ \text{a.s.}~P_{0}.  \label{eq3: proof}
\ee 
The proof follows combining (\ref{eq1: proof}), (\ref{eq2: proof}) and (\ref{eq3: proof}).
\end{Proof*}

\section{Collapsed Conditional of $\bk$} \label{appendix: posterior of k}
The two key steps in deriving the collapsed conditional of $\bk$ in Section \ref{sec: posterior computation} were (a) to obtain a closed form expression for $p(k_{j}\mid \bz_{j},\bw_{j})$ and then (b) to show that $p(\bk\mid \by,\bz,\bz^{\star},\blambda^{\star},\bpi^{\star}) = \prod_{j=1}^{q}p(k_{j}\mid \bz_{j}, \bw_{j})$. 
Part (b) follows easily by noting that the Markov blanket of $k_{j}$, after $\bpi_{\bk}$ are integrated out, comprises precisely $\bz_{j}$ and $\bw_{j}$.
We provide the technical details of the first step here. 
We use the generic $p_{0}$ to denote priors and hyper-priors. 
First, we note that integrating out $\bpi_{k_{j}}^{(j)}$ gives 
\be
&&\hspace{-1cm} p(\bz_{j} \mid \bw_{j}, k_{j})
= \int p(\bz_{j} \mid \bw_{j}, \bpi_{k_{j}}^{(j)}, k_{j}) p_{0}(\bpi_{k_{j}}^{(j)}) d\bpi_{k_{j}}^{(j)} 
= \prod_{r=1}^{C_{0}}\left\{\frac{\Gamma(k_{j}\gamma_{j})}{\{\Gamma(\gamma_{j})\}^{k_{j}}} \frac{\prod_{\ell=1}^{k_{j}}\Gamma\{\gamma_{j}+n_{j,r}(\ell)\}}{\Gamma(k_{j}\gamma_{j}+n_{j,r})} \right\}  \nonumber \\
&&= \prod_{r=1}^{C_{0}}\left\{\frac{1}{(k_{j}\gamma_{j})^{(n_{j,r})}} \prod_{\ell=1}^{\max\bz_{j,r}}\gamma_{j}^{(n_{j,r}(\ell))}  \right\} 
= \left\{\prod_{r=1}^{C_{0}}\prod_{\ell=1}^{\max\bz_{j,r}}\gamma_{j}^{(n_{j,r}(\ell))}  \right\}             \left\{\prod_{r=1}^{C_{0}}\frac{1}{(k_{j}\gamma_{j})^{(n_{j,r})}} \right\}, \label{eq: p(ZIW)}  
\ee 
where $x^{(m)}=x(x+1)\dots(x+m-1)$ with $x^{(0)}=1$, 
$\bz_{j,r} = \{z_{j,t}: w_{j,t}=r\}$, 
$n_{j,r} = \sum_{t=t^{\star}}^{T}1\{w_{j,t}=r\}$ denotes the frequency of the $r\th$ category of the $j\th$ predictor $w_{j}$ 
and $n_{j,r}(\ell)=\sum_{t=t^{\star}}^{T}1\{z_{j,t}=\ell,w_{j}=r\}$ denotes the number of allocation variables 
that are associated with the $r\th$ category of the $j\th$ predictor and are instantiated at $\ell$. 
Also, since $p(k_{j}\mid\bw_{j})=p_{0,j}(k_{j})$, we have  
\bse
&&\hspace{-1cm} p(\bz_{j}\mid \bw_{j}) 
= \sum_{k_{j}=\max_{r}\{\bz_{j,r}\}}^{C_{0}}     p(\bz_{j} \mid \bw_{j}, k_{j}) p_{0,j}(k_{j}) 
= \left\{\prod_{r=1}^{C_{0}}\prod_{\ell=1}^{\max\bz_{j,r}}\gamma_{j}^{(n_{j,r}(\ell))}  \right\}        U_{n_{j,1},\dots,n_{j,C_{0}}}(\max\bz_{j}),
\ese 
with $U_{n_{j,1},\dots,n_{j,C_{0}}}(z) = \sum_{k_{j}=z}^{C_{0}}  p_{0,j}(k_{j})   \prod_{r=1}^{C_{0}}\{\Gamma(k_{j}\gamma_{j})/\Gamma(k_{j}\gamma_{j}+n_{j,r})\}$.
This yields a closed form expression for $p(k_{j}\mid \bz_{j},\bw_{j})$ as
\be
&&\hspace{-1cm} p(k_{j}\mid \bz_{j}, \bw_{j}) = \frac{p(k_{j}\mid \bw_{j})~p(\bz_{j} \mid \bw_{j}, k_{j})}{p(\bz_{j}\mid \bw_{j})}  
= \frac{  p_{0,j}(k_{j})\prod_{r=1}^{C_{0}}\frac{\Gamma(k_{j}\gamma_{j})}{\Gamma(k_{j}\gamma_{j}+n_{j,r})}  }      {U_{n_{j,1},\dots,n_{j,C_{0}}}(\max\bz_{j})}, ~~~~~k_{j}=\max\bz_{j},\dots,C_{0}. \nonumber\\ 
\label{eq: p(k|Z,W) 2}
\ee
This completes the derivation. 

\end{appendix}

\vspace{1cm}
\baselineskip=14pt
\bibliographystyle{natbib}
\bibliography{HOMC}

\clearpage\pagebreak\newpage
\newgeometry{textheight=9in, textwidth=6.5in,}
\pagestyle{fancy}
\fancyhf{}
\rhead{\bfseries\thepage}
\lhead{\bfseries SUPPLEMENTARY MATERIALS}

\baselineskip 20pt
\begin{center}
{\LARGE{Supplementary Materials} 
for\\ {\bf Bayesian Nonparametric Modeling of Higher Order Markov Chains}}
\end{center}

\setcounter{equation}{0}
\setcounter{page}{1}
\setcounter{table}{1}
\setcounter{figure}{0}
\setcounter{section}{0}
\numberwithin{table}{section}
\renewcommand{\theequation}{S.\arabic{equation}}
\renewcommand{\thesubsection}{S.\arabic{section}.\arabic{subsection}}
\renewcommand{\thesection}{S.\arabic{section}}
\renewcommand{\thepage}{S.\arabic{page}}
\renewcommand{\thetable}{S.\arabic{table}}
\renewcommand{\thefigure}{S.\arabic{figure}}
\baselineskip=15pt

\vspace{-0.5cm}
\begin{center}
Abhra Sarkar and David B. Dunson\\
Department of Statistical Science, Duke University, Box 9025, Durham NC 27708-0251\\
abhra.sarkar@stat.duke.edu and dunson@duke.edu
\end{center}

\section{Prior Hyper-parameter $\varphi$}

\vspace{-0.5cm}
\begin{figure}[ht!]
\begin{center}
        \includegraphics[height=13.5cm,width=11cm, trim=4cm 6cm 3cm 5cm]{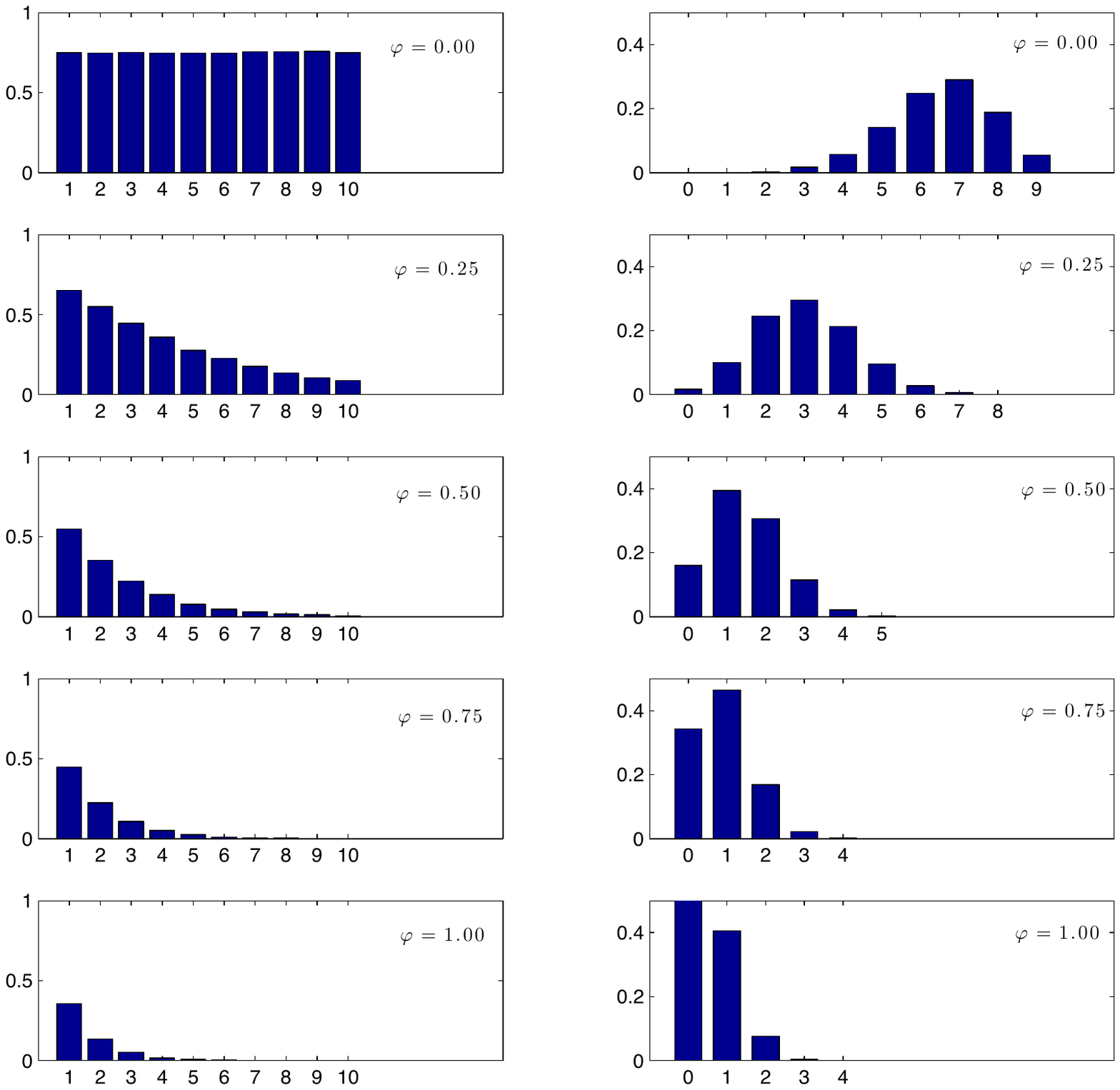}
        \caption{Induced prior probabilities of different lags to be included in the model ($k_{j}>1$, left panels)
        and the total number of important lags ($\sum_{j=1}^{q}1\{k_{j}>1\}$, right panels) 
        for the proposed conditional tensor factorization based Markov model with $C_{0}=4$ states and maximal order $q=10$ 
        for various values of $\varphi$ under the prior (\ref{eq: prior on k_{j}'s}) of the main paper.
        }
        \label{fig: varphi}
\end{center}
\end{figure}

\section{Approximate Sampler}  \label{sec: approximate sampler}

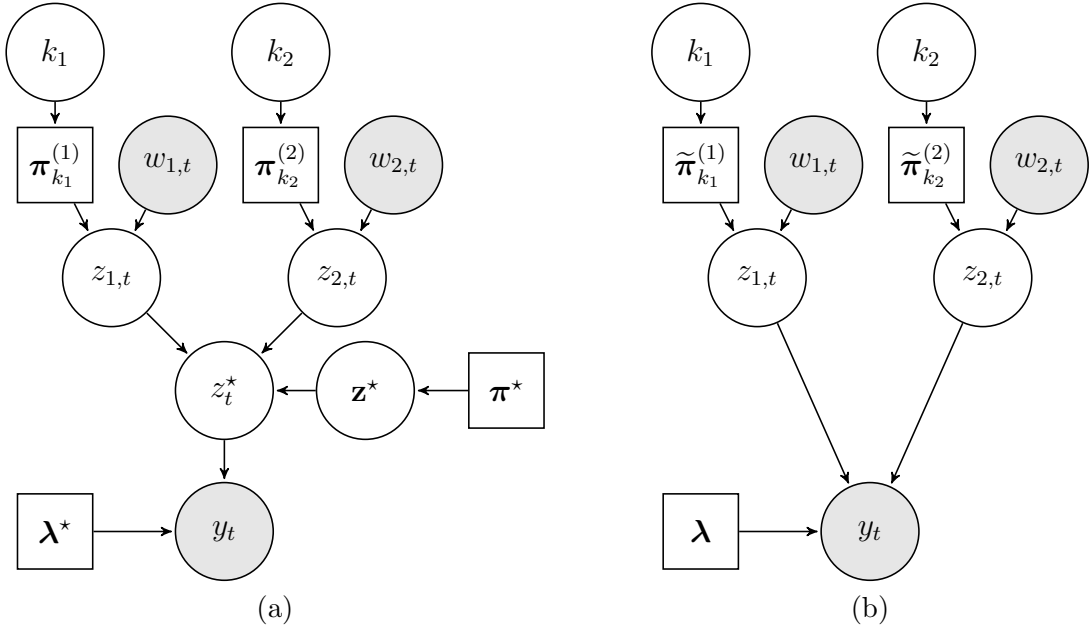
\begin{figure}[ht!]
\subfloat[][]
{
\centering
\begin{tikzpicture}[scale=1.5,->,>=stealth',shorten >=1pt,auto,node distance=2.8cm,semithick]

  \node[style={draw,circle}, minimum size=1.3cm] (k_1) at (0,1) {$k_{1}$};
  \node[style={draw,circle}, minimum size=1.3cm] (k_2) at (2,1) {$k_{2}$};

  \node[style={draw,rectangle}, minimum size=1cm] (pi_1) at (0,0) {$\bpi_{k_{1}}^{(1)}$};
  \node[style={draw,rectangle}, minimum size=1cm] (pi_2) at (2,0) {$\bpi_{k_{2}}^{(2)}$};

  \node[style={draw,circle,fill=gray!20}, minimum size=1.3cm] (w_1_t) at (1,0) {$w_{1,t}$};
  \node[style={draw,circle,fill=gray!20}, minimum size=1.3cm] (w_2_t) at (3,0) {$w_{2,t}$};

  \node[style={draw,circle}, minimum size=1.3cm] (z_1_t) at (0.5,-1) {$z_{1,t}$};
  \node[style={draw,circle}, minimum size=1.3cm] (z_2_t) at (2.5,-1) {$z_{2,t}$};

  \node[style={draw,circle}, minimum size=1.3cm] (z_t_star) at (1.5,-2) {$z_{t}^{\star}$};

  \node[style={draw,rectangle}, minimum size=1cm] (pi_star) at (4,-2) {$\bpi^{\star}$};

  \node[style={draw,circle}, minimum size=1.3cm] (z_star) at (2.75,-2) {$\bz^{\star}$};

  \node[style={draw,rectangle}, minimum size=1cm] (lambda_star) at (0,-3.25) {$\blambda^{\star}$};
  
  \node[style={draw,circle,fill=gray!20}, minimum size=1.3cm] (y_t) at (1.5,-3.25) {$y_{t}$};

  \path (k_1) edge (pi_1);
  \path (k_2) edge (pi_2);

  \path (pi_1) edge (z_1_t);
  \path (pi_2) edge (z_2_t);

  \path (z_1_t) edge (z_t_star);
  \path (z_2_t) edge (z_t_star);

  \path (pi_star) edge (z_star);
  
  \path (z_star) edge (z_t_star);

  \path (w_1_t) edge (z_1_t);
  \path (w_2_t) edge (z_2_t);

  \path (z_t_star) edge (y_t);
  \path (lambda_star) edge (y_t);

\end{tikzpicture}
\label{fig: graph detailed (a)}
}
\hspace{1cm}
\subfloat[][]
{
\centering
\begin{tikzpicture}[scale=1.5,->,>=stealth',shorten >=1pt,auto,node distance=2.8cm,semithick]

  \node[style={draw,circle}, minimum size=1.3cm] (k_1) at (0,1) {$k_{1}$};
  \node[style={draw,circle}, minimum size=1.3cm] (k_2) at (2,1) {$k_{2}$};

  \node[style={draw,rectangle}, minimum size=1cm] (pi_1) at (0,0) {$\wt\bpi_{k_{1}}^{(1)}$};
  \node[style={draw,rectangle}, minimum size=1cm] (pi_2) at (2,0) {$\wt\bpi_{k_{2}}^{(2)}$};

  \node[style={draw,circle,fill=gray!20}, minimum size=1.3cm] (w_1_t) at (1,0) {$w_{1,t}$};
  \node[style={draw,circle,fill=gray!20}, minimum size=1.3cm] (w_2_t) at (3,0) {$w_{2,t}$};

  \node[style={draw,circle}, minimum size=1.3cm] (z_1_t) at (0.5,-1) {$z_{1,t}$};
  \node[style={draw,circle}, minimum size=1.3cm] (z_2_t) at (2.5,-1) {$z_{2,t}$};


  \node[style={draw,rectangle}, minimum size=1cm] (lambda) at (0,-3.25) {$\blambda$};
  
  \node[style={draw,circle,fill=gray!20}, minimum size=1.3cm] (y_t) at (1.5,-3.25) {$y_{t}$};

  \path (k_1) edge (pi_1);
  \path (k_2) edge (pi_2);

  \path (pi_1) edge (z_1_t);
  \path (pi_2) edge (z_2_t);
  
  \path (w_1_t) edge (z_1_t);
  \path (w_2_t) edge (z_2_t);

  \path (z_1_t) edge (y_t);
  \path (z_2_t) edge (y_t);
  
  \path (lambda) edge (y_t);

\end{tikzpicture}
\label{fig: graph detailed (b)}
}
\caption{Graphical model depicting the dependency structure in a second order Markov chain $\{y_{t}\}$ for time point $t$.
(a) The proposed model implementing soft clustering of $z_{j,t} \sim \bpi_{k_{j}}^{(j)}(w_{j,t})$ with $\bpi_{k_{j}}^{(j)}(w_{j,t}) \sim \Dir(\gamma_{j},\dots,\gamma_{j})$ for all $j, w_{j,t}$, and $\blambda_{h_{1},\dots,h_{q}} \sim \sum_{\ell=1}^{\infty}\pi_{\ell}^{\star}\blambda_{\ell}^{\star}$ independently for all $(h_{1},\dots,h_{q})$.  
(b) An approximation of the proposed model implementing hard clustering of $z_{j,t} \sim \wt\bpi_{k_{j}}^{(j)}(w_{j,t})$ with $\wt\pi_{h_{j}}^{(j)}(w_{j,t}) \in \{0,1\}$ for all $j,h_{j}$ and $w_{j,t}$, and $\blambda_{h_{1},\dots,h_{q}} \sim \Dir(\alpha,\dots,\alpha)$ independently for all $(h_{1},\dots,h_{q})$. 
This approximation forms the basis of the approximate sampler described in Section \ref{sec: approximate sampler} of the Supplementary Materials.  
}
\end{figure}

This section describes the approximate sampler used to determine the starting values of the latent class allocation variables $\bz$ for the MCMC sampler described in Section \ref{sec: posterior computation} of the main paper. 

Given a model indexed by $\bk=\{k_{1},\dots,k_{q}\}$, the levels of $w_{j}$ are partitioned into $k_{j}$ clusters $\{\C_{j,r}: r=1,\dots,k_{j}\}$ with each cluster $\C_{j,r}$ assumed to correspond to its own latent class $h_{j}=r$. 
With independent Dirichlet priors on the mixture kernels $\blambda_{h_{1},\dots,h_{q}} \sim \Dir(\alpha,\dots,\alpha)$ marginalized out, 
the likelihood conditional on the cluster configurations $\C=\{\C_{j,r}: j=1,\dots,q, r=1,\dots,k_{j}\}$ is given by 
\be
p(\by \mid \C) = \prod_{(h_{1},\dots,h_{q})}  \frac{\beta\{\alpha+n_{h_{1},\dots,h_{q}}(1), \dots, \alpha+n_{h_{1},\dots,h_{q}}(C_{0})\}}{\beta(\alpha,\dots,\alpha)},
\ee
\vspace{-3ex}\\
where $n_{h_{1},\dots,h_{q}}(y) = \sum_{t=t^{\star}}^{T}1\{y_{t}=y,w_{1,t}\in\C_{1,h_{1}},\dots,w_{q,t}\in\C_{m,h_{q}}\}$.
Given the current model indexed by $\bk=\{k_{1},\dots,k_{q}\}$ and clusters $\C=\{\C_{j,r}: j=1,\dots,q, r=1,\dots,k_{j}\}$, we do the following for $j=1,\dots,q$.
\begin{enumerate}
\item
If $k_{j}<C_{0}$, we propose to increase $k_{j}$ to $(k_{j}+1)$. 
If $k_{j}>1$, we propose to decrease $k_{j}$ to $(k_{j}-1)$.
For $1<k_{j}<C_{0}$, the moves are proposed with equal probability. 
For $k_{j}=1$, the increase move is selected with probability $1$.
For $k_{j}=C_{0}$, the decrease move is selected with probability $1$. 
\item
If an increase move is proposed, we randomly split a cluster of $w_{j}$ into two clusters. 
We accept this move with acceptance rate based on the approximated marginal likelihood.
\item 
If a decrease move is proposed, we randomly merge two clusters of $w_{j}$ into a single cluster. 
We accept this move with acceptance rate based on the approximated marginal likelihood.  
\end{enumerate}
The latent class allocation variables $\bz$ are initialized at the cluster allocation variables returned by the approximate sampler after $100$ iterations.

\section{MCMC Diagnostics} \label{sec: mcmc diagnostics}

\begin{figure}[ht!]
\begin{center}
	\includegraphics[height=14cm,width=15cm, trim=0cm 0cm 0cm 0cm]{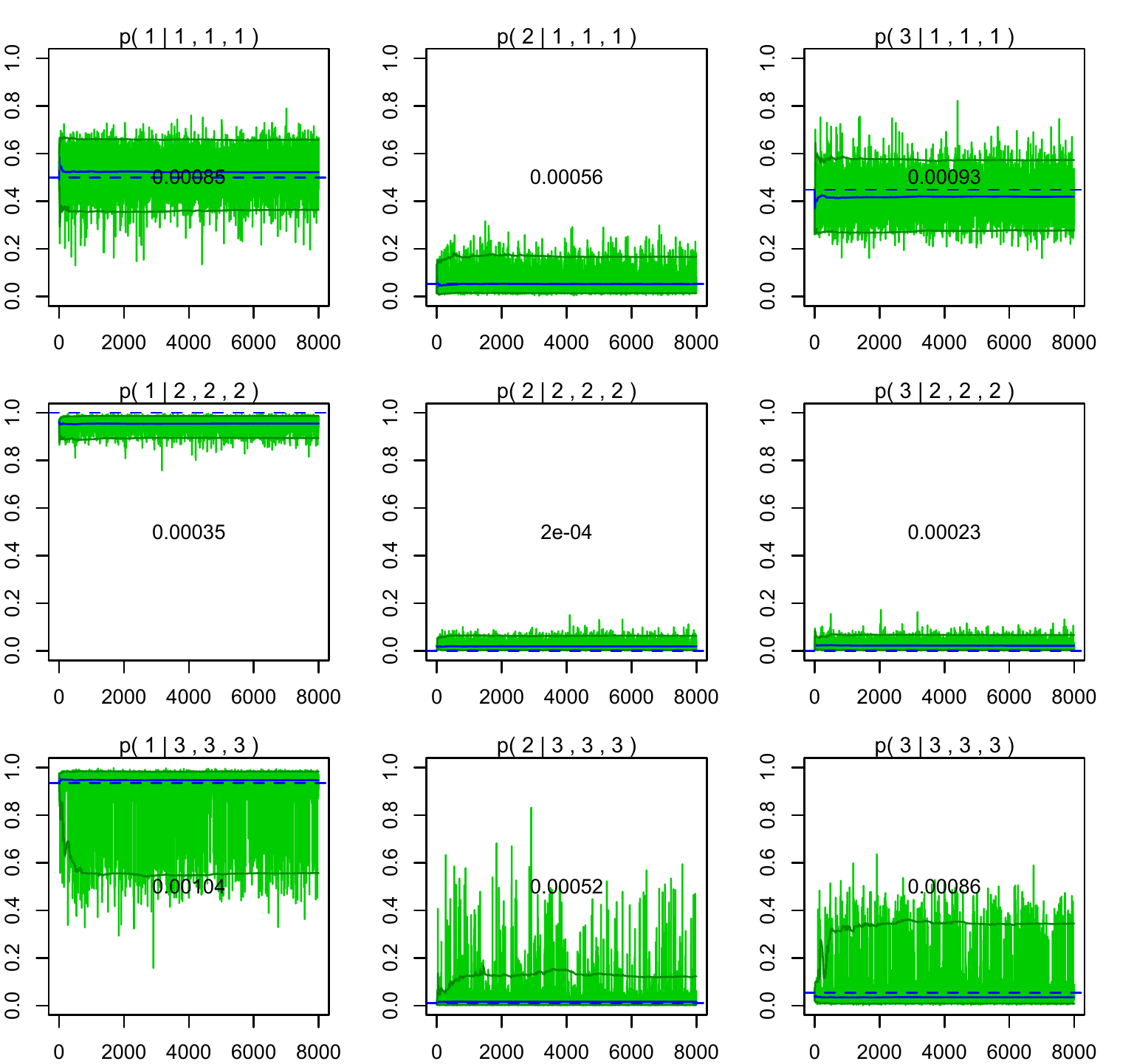}
	\caption{Trace plots of some transition probabilities for the case (G) with $C_{0}=3$ categories, true important lags $\{y_{t-1}, y_{t-4}, y_{t-8}\}$ and sample size $T = 500$ 
	for the data set with the median classification error rate in the simulation experiments. 
	In each panel, the solid blue line shows the running mean and the horizontal dashed blue line shows the corresponding true value. 
        The darker green lines show the 5\% and 95\% running quantiles. 
        The number at the middle of each panel shows the Monte Carlo standard error estimated by batch means analysis with batch length 100. 
        See Section \ref{sec: simulation experiments} of the main paper and Section \ref{sec: mcmc diagnostics} of the Supplementary Materials for additional details.  
        } 
        \label{fig: running mean sim}
\end{center}
\end{figure}

Figure \ref{fig: running mean sim} shows some additional MCMC diagnostics (Cowes and Carlin, 1996; Flegal and Jones, 2011) based on thinned samples 
for the case (G) with $C_{0}=3$ categories, true important lags $\{y_{t-1}, y_{t-4}, y_{t-8}\}$ and sample size $T = 500$ 
for the data set with the median classification error rate in the simulation experiments. 
Our model accommodates uncertainty in the set of important lags. This set may vary from one MCMC iteration to another. 
To draw the trace plot for $p(y \mid i_{1},i_{4},i_{8})$ accommodating variable lag sets, 
where $\{i_{1},i_{4},i_{8}\}$ denotes a specific value of $\{y_{t-1},y_{t-4},y_{t-8}\}$, 
we first identified a $t_{0}$ from $\{(T+1),\dots,N\}$ such that $\{y_{t_{0}-1},y_{t_{0}-4},y_{t_{0}-8}\}=\{i_{1},i_{4},i_{8}\}$. 
The trace plot for $p(y \mid i_{1},i_{4},i_{8})$ is then based on estimates of $p(y \mid y_{t_{0}-1},y_{t_{0}-2},\dots,y_{t_{0}-10})$ for different MCMC iterations.
Our experiments with different $t_{0}$'s with same $\{y_{t_{0}-1},y_{t_{0}-4},y_{t_{0}-8}\}=\{i_{1},i_{4},i_{8}\}$ produced very similar results. 
As Figure \ref{fig: running mean sim} shows, the running means and quantiles are very stable, Monte Carlo standard errors were small, 
and there is good agreement between the truth and the running posterior means.

\begin{figure}[ht!]
\begin{center}
	\includegraphics[height=14cm,width=15cm, trim=0cm 0cm 0cm 0cm]{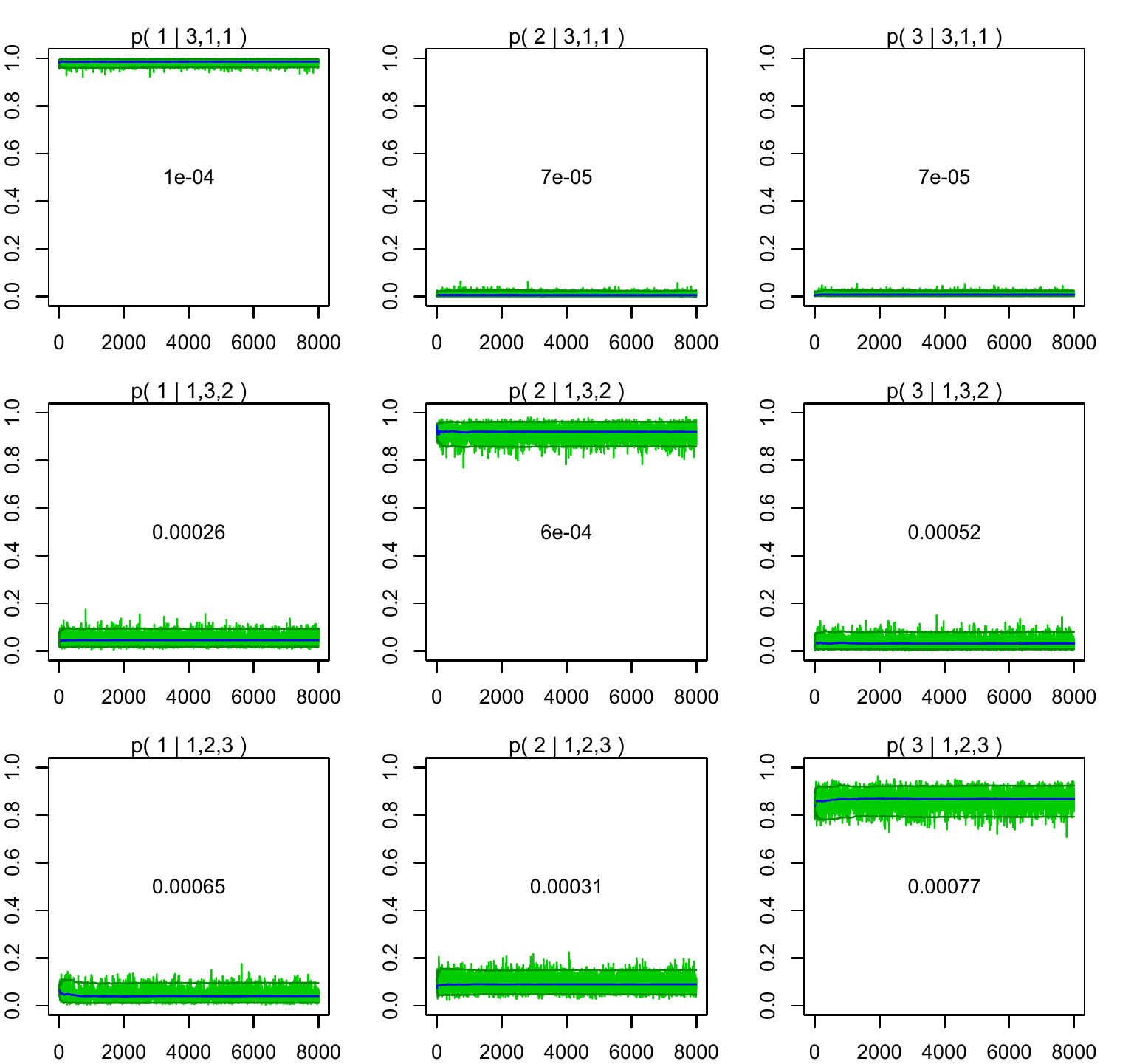}
	\caption{Trace plots of some transition probabilities for the wood pewee data set. 
        In each panel, the solid blue line shows the running mean. 
        The darker green lines show the 5\% and 95\% running quantiles. 
        The number at the middle of each panel shows the Monte Carlo standard error estimated by batch means analysis with batch length 100. 
        See Section \ref{sec: simulation experiments} of the main paper and Section \ref{sec: mcmc diagnostics} of the Supplementary Materials for additional details.  
        } 
        \label{fig: running mean wood pewee}
\end{center}
\end{figure}

Figure \ref{fig: running mean wood pewee} shows similar diagnostic plots for the MCMC output for the wood pewee data set analyzed in Section \ref{sec: wood pewee} of the main paper. 
In this case the truth is unknown. 
Following the discussion in Section \ref{sec: wood pewee}, we assumed $\{y_{t-1},y_{t-2},y_{t-4}\}$ to be the set of important lags. 
The trace plot for $p(y \mid i_{1},i_{2},i_{4})$ is thus based on the estimates of $p(y \mid y_{t_{0}-1},\dots,y_{t_{0}-10})$ for different MCMC iterations
for some $t_{0}$ from $\{T+1,\dots,N\}$ such that $\{y_{t_{0}-1},y_{t_{0}-2},y_{t_{0}-4}\}=\{i_{1},i_{2},i_{4}\}$. 
The running means and quantiles are again very stable with small Monte Carlo standard errors and the estimated posterior means agree well with our empirical expectations.

In all examples, the quantiles can be used to construct 90\% posterior probability regions. 
Since the transition probabilities have variances uniformly bounded above by $1/4$, 
Monte Carlo standard errors in the final posterior mean estimates have a conservative uniform upper bound of $1/(2\sqrt{8000})\approx 0.0056$.

\section{Analysis of Human Preproglucagon Gene Data Set} \label{sec: human gene data set}
In this section, we present an analysis of a DNA sequence found in the human preproglucagon gene (Bell \etal,1983). 
There are 1752 data points and four states A, C, G and T. 
\cite{avery_henderson:1999} and Besag and Mondal (2013) analyzed the data set focusing their attention on Markov models of up to third order, 
using asymptotic $\chi^{2}$ tests and simulation based exact tests, respectively, to assess fit. 
The $\chi^{2}$ test of a first order Markov chain against a second order alternative led to rejection of the null hypothesis at the level 0.019, 
and the test of a second order null versus a third order alternative produced a p-value of 0.34, 
whereas the corresponding simulation based tests produced p-values of 0.028 and 0.44, respectively, 
providing evidence that a second order model gives the best fit to the data set among the candidate models. 

Figure \ref{fig: Application 3: Human Gene Data Set, q=3} summarizes the results produced by the proposed conditional tensor factorization approach with maximal order $q=3$ applied to the first 1000 data points. 
Due to minor mixing issues, in this case we ran the MCMC algorithm for 5 million iterations (to be conservative) with the initial 2 millions discard as burn-in.  
The estimated posterior probabilities of first, second and third order Markov models were approximately $0$ (the MCMC chain never visited this model), $0.43$ and $0.57$, respectively. 
With a posterior odds of $\infty$ for a second order model against a first order model and a posterior odds of $1.33$ for a third order model against a second order model, 
the results were in general agreement with the frequentist analyses.  

The proposed conditional tensor factorization based approach enables us to test for serial dependencies of much higher orders. With the maximal order set at $q=10$, 
the posterior probability of the model being of maximal order $7$ was estimated to be approximately $0.94$. 
The MCMC algorithm never visited a Markov model of maximal order $2$. 
Figure \ref{fig: Application 3: Human Gene Data Set, q=10} summarizes the results.  Markov models are widely used for nucleotide sequences, and hence the ability to fit more realistic models containing higher order dependence is of substantial importance in this application area.
\vspace{-0.5cm}

\begin{figure}[ht]
\begin{center}
        \includegraphics[height=10cm,width=11cm, trim=3cm 4.5cm 3cm 5cm]{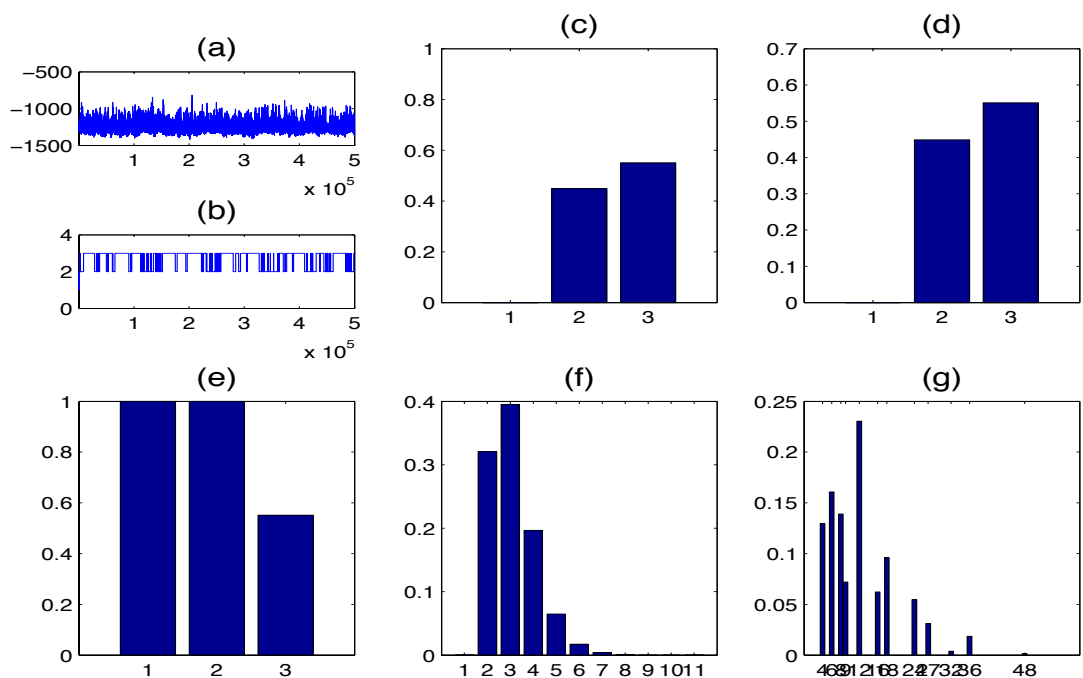}
        \caption{Results for the human preproglucagon gene data set with the maximal order set at $q=3$. 
         (a) trace plot of marginal likelihood $p(\by\mid \bz)$; 
         (b) trace plot of the number of important lags; 
         (c) relative frequency distribution of the maximal order; 
         (d) relative frequency distribution of the number of important lags; 
         (e) inclusion proportions of different lags; 
         (f) relative frequency distribution of the number of clusters of the probability kernels $\blambda_{h_{1},\dots,h_{q}}$; and   
         (g) relative frequency distribution of $\prod_{j=1}^{q}k_{j}$, the number of possible combinations of $(h_{1},\dots,h_{q})$. 
        Panels (c)-(g) are based on thinned samples after burn-in. 
        }
        \label{fig: Application 3: Human Gene Data Set, q=3}
\end{center}
\end{figure}

\begin{figure}[ht]
\begin{center}
        \includegraphics[height=10cm,width=11cm, trim=3cm 4.5cm 3cm 5cm]{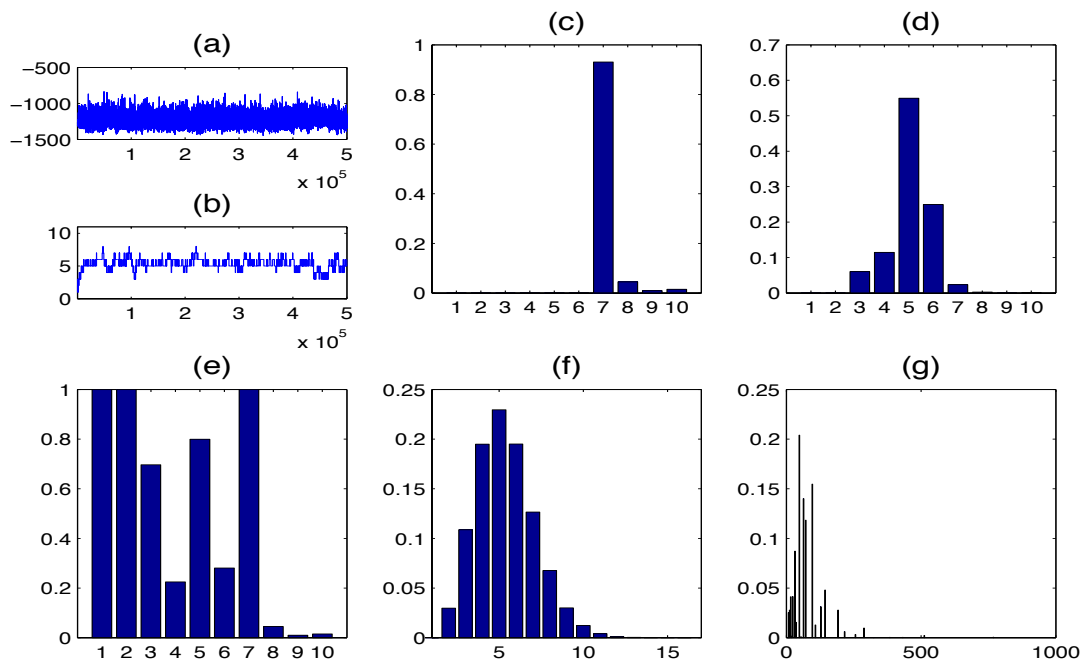}
        \caption{Results for the human preproglucagon gene data set with the maximal order set at $q=10$. 
         (a) trace plot of marginal likelihood $p(\by\mid \bz)$; 
         (b) trace plot of the number of important lags; 
         (c) relative frequency distribution of the maximal order; 
         (d) relative frequency distribution of the number of important lags; 
         (e) inclusion proportions of different lags; 
         (f) relative frequency distribution of the number of clusters of the probability kernels $\blambda_{h_{1},\dots,h_{q}}$; and   
         (g) relative frequency distribution of $\prod_{j=1}^{q}k_{j}$, the number of possible combinations of $(h_{1},\dots,h_{q})$. 
        Panels (c)-(g) are based on thinned samples after burn-in. 
        }
        \label{fig: Application 3: Human Gene Data Set, q=10}
\end{center}
\end{figure}

\section{Results Produced by VLMC and RFMC}  \label{sec: vlmc & rfmc summary}
\begin{figure}[ht] 
\begin{minipage}[c][11cm][t]{.5\textwidth}
  \vspace*{\fill}
  \centering
  \includegraphics[height=10cm,width=6cm, trim=2cm 0cm 2cm 0cm]{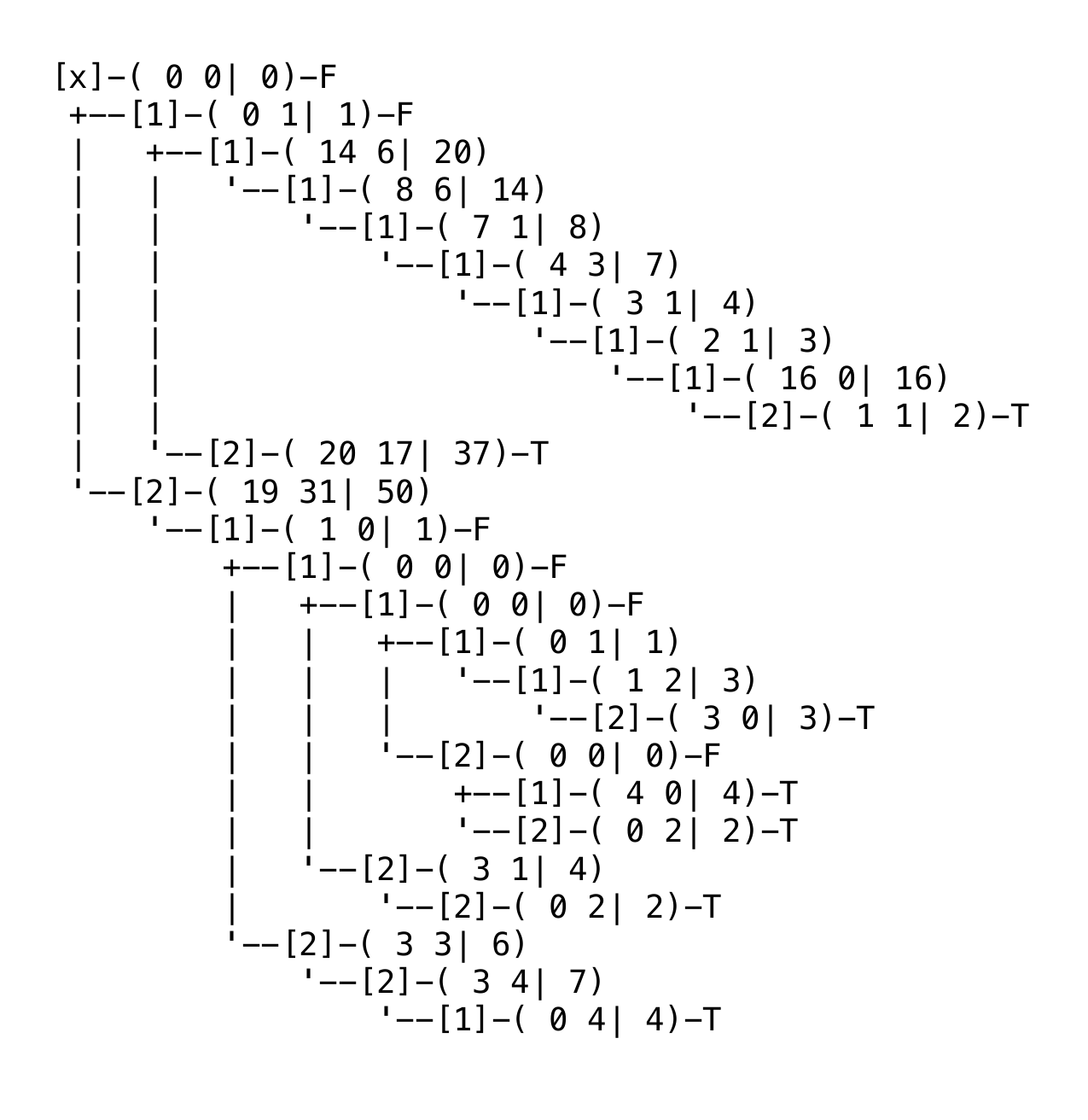}
  \captionof{subfigure}{Epileptic Seizure Data Set}
  \label{fig:2:test1}
\end{minipage}%
\begin{minipage}[c][11cm][t]{.5\textwidth}
  \vspace*{\fill}
  \centering
  \includegraphics[width=5cm,height=4.5cm, trim=1cm 0cm 1cm 0cm]{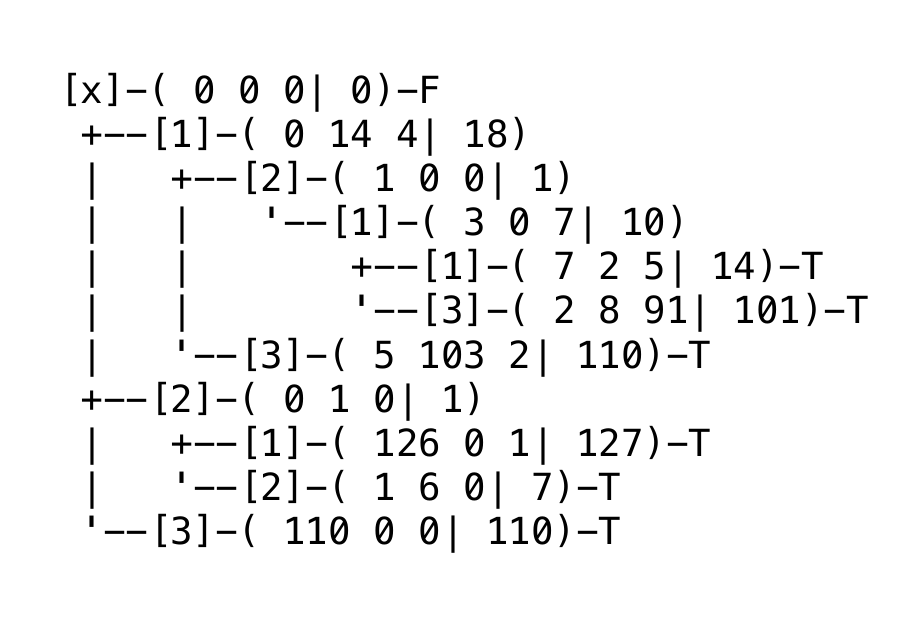}
  \captionof{subfigure}{Song of the Wood Pewee Data Set}
  \par\vfill
  \includegraphics[width=5cm,height=4.5cm, trim=1cm 0cm 1cm 0cm]{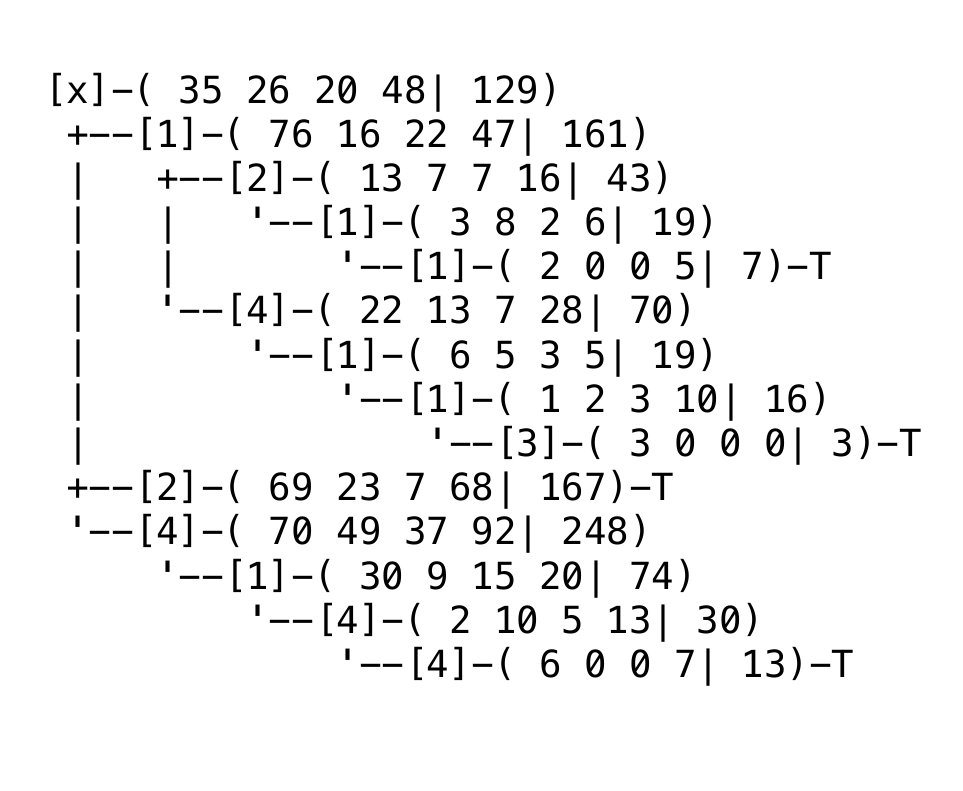}
  \captionof{subfigure}{Human Preproglucagon Gene Data Set}
\end{minipage}
\caption{Context trees produced by the VLMC method for the data sets discussed in Section \ref{sec: applications} of the main paper and Section \ref{sec: human gene data set} of the Supplementary Materials.}
\label{fig: VLMC context trees}
\end{figure}

Figure \ref{fig: VLMC context trees} shows the context trees \citep[see][]{machler_buhlmann:2004} summarizing the serial dependence structures and transition probabilities estimated by the VLMC method, 
as implemented by the VLMC package in R, 
applied to the real data sets discussed in Section \ref{sec: applications} of the main paper and 
Section \ref{sec: human gene data set} of the Supplementary Materials. The pruning parameter was selected by AIC criterion. 

\begin{figure}[ht] 
\begin{minipage}[c][6cm][t]{.3\textwidth}
  \par\vfill
  \centering
  \includegraphics[width=5.5cm,height=5cm, trim=0cm 0cm 0cm 2cm,clip=true]{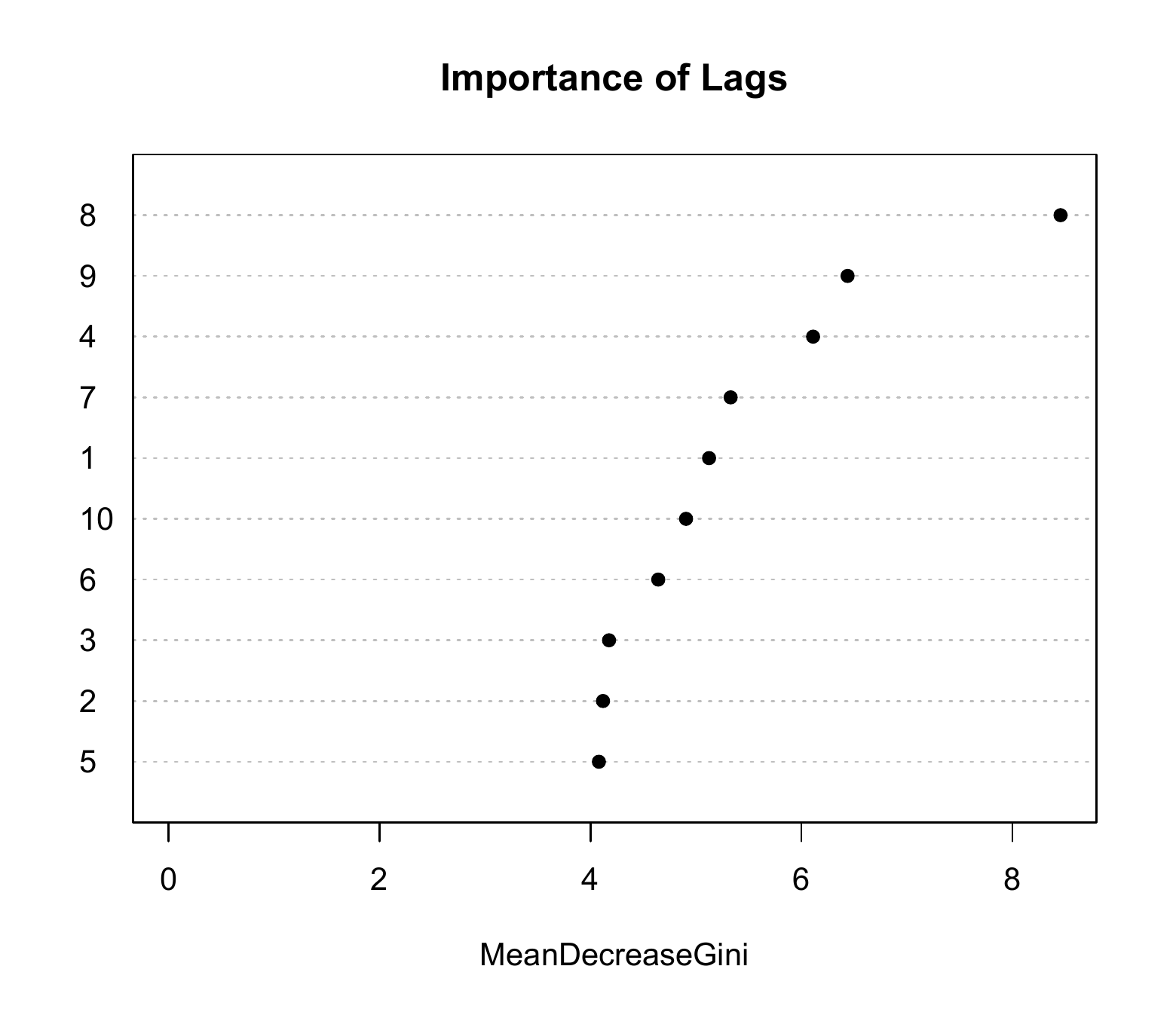}
  \captionof{subfigure}{Epileptic Seizure}
  \label{fig:2:test1}
\end{minipage}%
\begin{minipage}[c][6cm][t]{.3\textwidth}
  \par\vfill
  \centering
  \includegraphics[width=5.5cm,height=5cm, trim=0cm 0cm 0cm 2cm,clip=true]{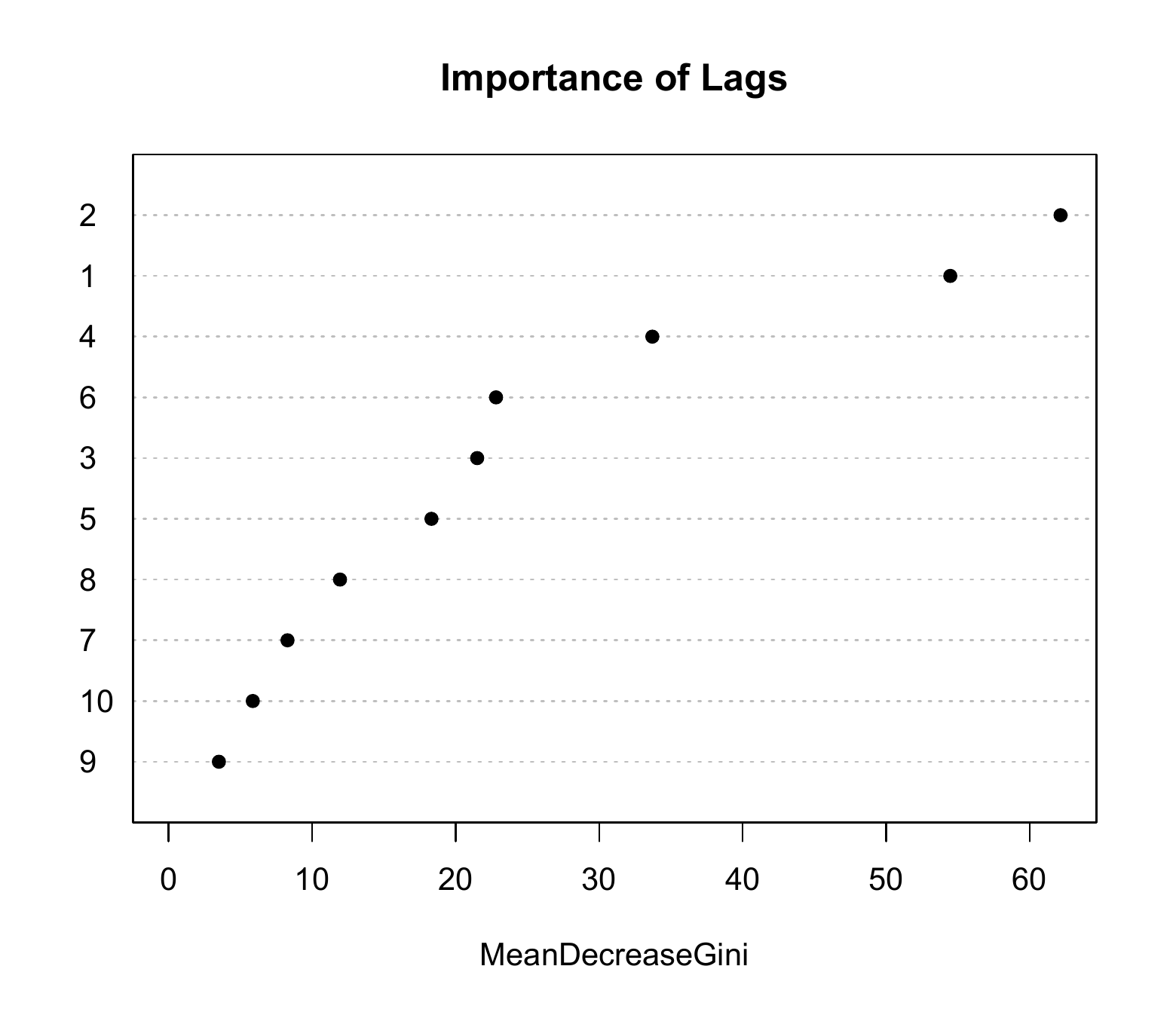}
  \captionof{subfigure}{Song of the Wood Pewee}
\end{minipage}
\begin{minipage}[c][6cm][t]{.3\textwidth}
  \par\vfill
  \centering
  \includegraphics[width=5.5cm,height=5cm, trim=0cm 0cm 0cm 2cm,clip=true]{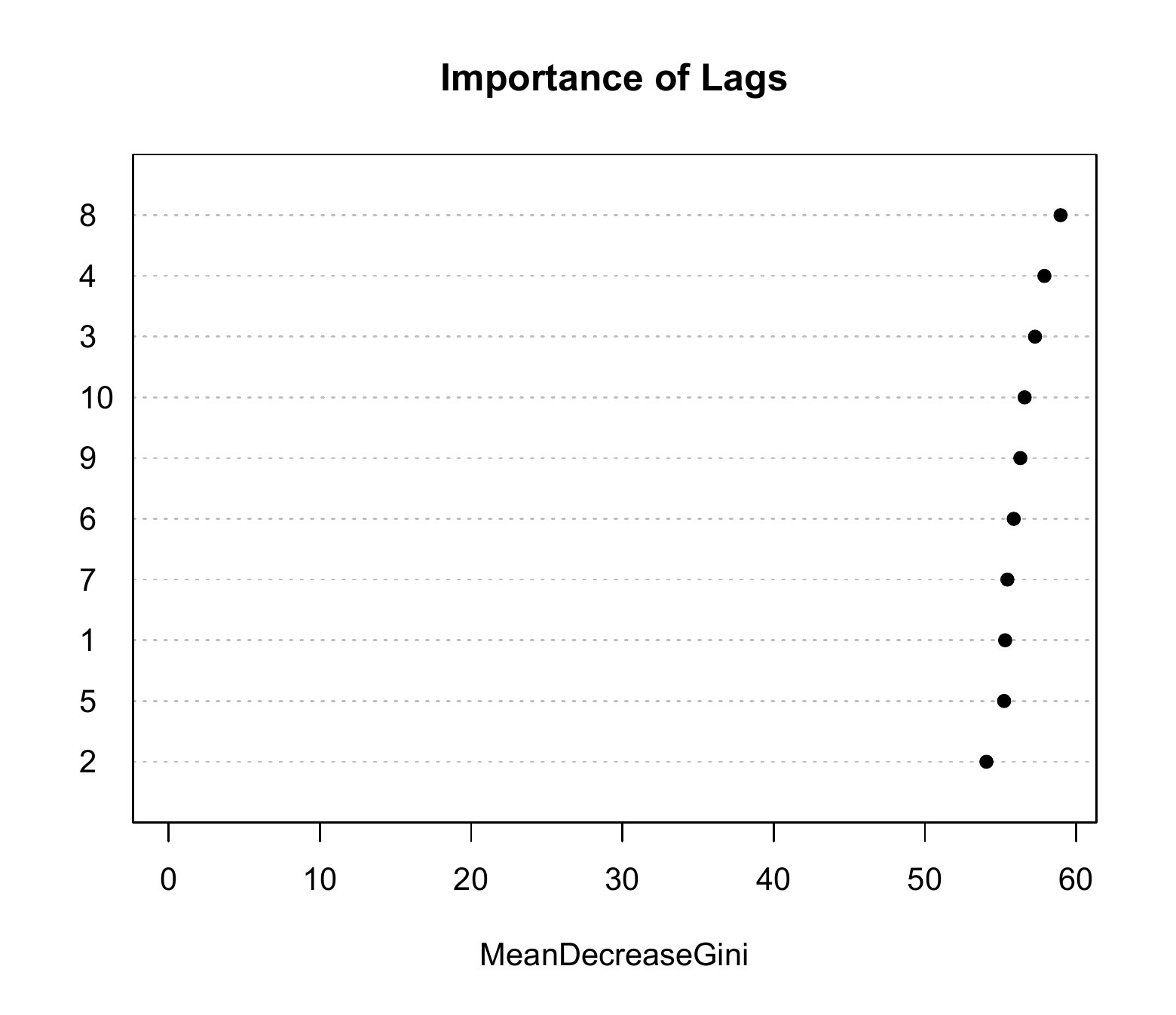}
  \captionof{subfigure}{Human Gene}
  \label{fig:2:test1}
\end{minipage}%
\vspace{20pt}
\caption{Plots showing the relative importance of different lags as estimated by the RFMC method for the data sets discussed in Section \ref{sec: applications} of the main paper and Section \ref{sec: human gene data set} of the Supplementary Materials.}
\label{fig: RFMC lag imp plots}
\end{figure}

Figure \ref{fig: RFMC lag imp plots} shows the relative importance of different lags estimated by the RFMC method, 
as implemented by the radomForest package in R, 
applied to the real data sets discussed in Section \ref{sec: applications} of the main paper and Section \ref{sec: human gene data set} of the Supplementary Materials. 

For the epileptic seizure data set, the wood pewee data set and the human gene data set, the VLMC method estimated Markov chains of maximal orders $9$, $4$ and $4$, respectively.  
For the seizure data set, the entire sequence consisted of only $204$ data points. 
With the first $200$ data points used to fit the models, there were not enough additional observations to evaluate prediction performances. 
For the wood pewee data set and the human gene data set, we used the first $500$ and $1000$ data points to fit the models and the following $500$ observations to evaluate one-step ahead prediction performances. 
For the wood pewee data set, classification error rates for our proposed conditional tensor factorization based approach, VLMC and RFMC were $0.021$, $0.024$ and $0.026$, respectively. 
For the human gene data set, classification error rates for our proposed conditional tensor factorization based approach, VLMC and RFMC were $0.64$, $0.66$ and $0.66$, respectively. 

While the maximal orders and the classification error rates estimated by the two competing methods are in general agreement,  
the VLMC method, with a top-down tree based mechanism to model serial dependencies, 
could not detect gaps in the set of important lags for the first two data sets, which were suggested by the proposed conditional tensor factorization approach.  Such gaps seem to be a common feature of many data sets, which is commonly obscured by existing statistical methods.

\newpage
\section*{Additional References}
\refmark
Bell, I. G., Sanchez-Pescador, R. Laybourn, P. J., and Najarian, R. C. (1983). Exon
duplication and divergence in the human preproglucagon gene. \emph{Nature}, {\bf 304}, 368-371.

\refmark
Cowes, M. K. and Carlin, B. P. (1996). Markov chain Monte Carlo convergence diagnostics: A comparative review. 
\JASA, {\bf 91}, 883-904.

\refmark
Flegal, M. and Jones, G. (2011). 
Implementing MCMC: Estimating with confidence. In S. Brooks, A. Gelman, G. Jones, and X. Meng, editors, 
\emph{Handbook of Markov chain Monte Carlo}, pages 175-197. Chapman \& Hall/CRC Press.


\end{document}